\documentclass{elsarticle}

\usepackage{hyperref}
\usepackage{color}
\usepackage{amsmath}
\usepackage{amsfonts}
\usepackage{amssymb}
\usepackage{graphicx}
\usepackage{fullpage}
\usepackage{xcolor}
\usepackage{setspace}
\usepackage{amsthm}
\usepackage{algorithm}
\usepackage{algorithmic}
\usepackage{fancybox}
\usepackage{fancyhdr}
\usepackage{bbm}
\usepackage{multirow}
\usepackage{subfig}
\usepackage{natbib}

\DeclareMathOperator*{\Cov}{Cov}
\DeclareMathOperator*{\Var}{Var}

\newcommand{\indd}{\mathbbm{1}}
\newcommand{\ensemble}[1]{ \{ #1 \} }
\newcommand{\R}{\mathbb{R}}
\newcommand{\E}{\mathbb{E}}
\newcommand{\Proba}{\mathbb{P}}
\newcommand{\Bcal}{\mathcal{B}}
\newcommand{\Dn}{\mathcal{D}_n}
\newcommand{\Dnbar}{\bar{\mathcal{D}}_n}

\theoremstyle{plain}

\newtheorem{prop}{Proposition}
\newtheorem{cor}{Corollary}

\newtheorem*{remark}{Remark}

\theoremstyle{definition}

\journal{Journal of Computational Statistics $\&$ Data Analysis}

\biboptions{authoryear}

\begin{document}

\begin{frontmatter}

\title{Grouped variable importance with random forests and application to multiple functional data analysis}



\author[Safety,upmc]{Baptiste Gregorutti\corref{mycorrespondingauthor}}
\cortext[mycorrespondingauthor]{Corresponding author}
\ead{baptiste.gregorutti@safety-line.fr}

\author[upmc]{Bertrand Michel}
\ead{bertrand.michel@upmc.fr}

\author[upmc]{Philippe Saint-Pierre}
\ead{philippe.saint$\_$pierre@upmc.fr}

\address[Safety]{Safety Line, 15 rue Jean-Baptiste Berlier, 75013 Paris, France}
\address[upmc]{Laboratoire de Statistique Th\'eorique et Appliqu\'ee, Sorbonne Universit\'es, UPMC Univ Paris 06, F-75005, Paris, France}

\begin{abstract}
The selection of grouped variables using the random
forest algorithm is considered. First a new importance measure adapted for groups
of variables is proposed. Theoretical insights into this criterion are given for additive regression
models. Second, an original method for selecting
functional variables based on the grouped variable importance measure is developed. Using a
wavelet basis, it is proposed to regroup all of the wavelet coefficients for a given functional variable and use a
wrapper selection algorithm with these groups. Various
other groupings which take advantage of the frequency and time localization of the
wavelet basis are proposed. An extensive simulation study is performed to illustrate
the use of the grouped importance measure in this context. The method is applied
to a real life problem coming from aviation safety.

\end{abstract}


\end{frontmatter}

\section{Introduction}

In the high dimensional setting,  identification of the most relevant variables has been the subject of much
research during the last two decades~\citep{guyon2003introduction}. For linear regression, the  lasso
method \citep{tib96} is widely used. Many variable selection procedures have also been proposed for non linear methods.
In the context of random forests \citep{rf:B01}, it has been shown that the permutation importance measure
is an efficient tool for selecting variables \citep{rf:DA05,rf:GPT10,rf:G+14}.

In many situations such as medical studies and genetics, groups of variables can be clearly identified and it is of
interest to select  groups of variables rather than to select  them individually \citep{he2010stable}. Indeed,
interpretation of the model may be improved along with the prediction accuracy by grouping the variables according to
\emph{a priori} knowledge about
the data. Furthermore, grouping   variables can be seen  as a solution to stabilize
variable selection methods. In the linear setting, and more particularly for
 linear regression, the group lasso has been developed to deal with groups of variables, see for
instance \citet{lasso:YL06}. Group variable selection has also been proposed for kernel methods
\citep{zhang2008variable} and neural networks \citep{fs:CP08}. As far as we know, this problem has not been studied for
the random forest algorithm introduced by
\citet{rf:B01}. In this  paper, we adapt the permutation importance measure for groups of
variables in order to select groups of variables in the context of random forests.

The first contribution of this paper is a theoretical analysis of the  grouped variable importance measure.
Generally speaking, the
grouped variable importance does not reduce to the sum of the individual importances and may even be quite unrelated to it. However, in more specific models such as additive regression ones, we derive exact decompositions of
the grouped variable importance measure.

The second contribution of this work is an original method for selecting functional
variables based on the grouped variable importance measure. Functional Data Analysis (FDA) is a field in statistics that analyzes data indexed by a continuum. In our case, we consider data providing information about curves varying over time \citep{fda:RS05,fda:FV06,fda:F11}. One
standard approach in FDA  consists in projecting the functional variables onto a finite dimensional space
spanned by a  functional basis. Classical bases in this context are splines, Fourier, wavelets or Karhunen-Lo\`eve expansions, for
instance. Most of the papers about regression and classification methods for functional data consider only one
functional predictor; references include
\citet{fda:C+99,fda:CFS03,fda:R+06,fda:CH06} for
  linear regression methods,  \citet{fda:A+06,fda:A+09} for logistic regression methods,  \citet{anovaFd} for ANOVA problem, \citet{fda:BBW05,fda:FT06} for $k$-NN algorithms and  \citet{fda:RV06,fda:RV08} for SVM classification.
The multiple FDA problem, where $p$ functional variables are observed, has been less studied. Recently,
\citet{fda:MK11} and \citet{fda:FJ13} have proposed solutions to the linear regression problem with lasso-like
penalties. The logistic regression case has been studied by \citet{Matsui2014176}. Classification based on several
functional variables has also been considered  using the CART algorithm \citep{fda:PT06} and SVM
\citep{svm:YYS05,svm:YS06}.

We propose a new approach for multiple FDA using random forests and the grouped variable
importance measure. Indeed,  various groups of basis coefficients can be proposed for a given  functional
decomposition. For instance, one can choose to regroup all
coefficients of a given functional variable. In this case, the selection of a group of coefficients
corresponds to the selection of a functional variable. Various other groupings are proposed  for wavelet
decompositions. For a given family of groups, we adapt the recursive feature elimination algorithm \citep{svm:guyon2002}
which is particularly efficient when  predictors are strongly correlated \citep{rf:G+14}. In the context of random
forests, this backward-like selection algorithm is guided by the grouped variable importance. Note
that by
regrouping the coefficients, the computational cost of the algorithm is drastically reduced compared to a backward
strategy that would eliminate only one coefficient at each step.

 An extensive  simulation study illustrates the application of the grouped importance measure for FDA. The method is then
applied to a real life problem coming from aviation safety. The aim of this study is to explain  and predict landing
distances. We  select the most relevant flight parameters regarding the risk of long landings, which is a major issue
for airlines.

The group permutation importance measure is introduced in Section~\ref{sec:RFgrimp}.
Section~\ref{sec:contrib} deals with multiple FDA using random forests and the grouped variable importance measure.
The application to flight data analysis is presented in Section~\ref{sec:appli}. Note that additional experiments about
the grouped variable importance are given in \ref{sub:ExpPIGV}. In order to speed up the algorithm,
the dimension of the data can be reduced in a preprocessing step.
In \ref{sec:dimRed},  we propose a modified version of a well-known shrinkage method \citep{wave:DJ94}  that
simultaneously shrinks to zero the coefficients of the observed curves of a functional variable.

\section{The grouped variable importance measure}
\label{sec:RFgrimp}

Let $Y$ be a random variable in $\R$ and $\mathbf{X} ^\top= (X_1, \dots, X_p)$ a random vector in $ \R^p$. We denote by  $f(\mathbf x)=\E[Y|\mathbf X=\mathbf x]$  the regression function.  Let $\Var(\mathbf X)$ and $\Cov(\mathbf X)$ denote the variance and  variance-covariance matrices of $\mathbf X$.

The permutation importance introduced by \citet{rf:B01} measures the accuracy of each variable $X_j$ for predicting $Y$. It is based on the elementary property that the quadratic risk $ \E \left[ \left( Y - f(\mathbf X )  \right)^2 \right]$ is the minimum error for predicting $Y$ knowing $\mathbf X$. The formal definition of the variable importance measure of $X_j$ is:
\begin{equation}
\mathcal I(X_j) := \E \left[ \left( Y - f(\mathbf X_{(j)} ) \right)^2 \right] - \E \left[ \left( Y - f(\mathbf X )  \right)^2 \right],
\label{eq:rawImp}
\end{equation}
where $\mathbf X_{(j)} = (X_1, \ldots, X_j', \ldots, X_p)^\top$ is a random vector such that $X_j'$ is an independent replicate of $X_j$ which is also independent of $Y$ and of all  other predictors. This criterion evaluates the increase of the prediction error after breaking the link between the variable $X_j$ and the outcome $Y$ (see \citet{rf:Z+12} for instance).

In this paper, we extend the permutation importance to groups of
variables. Let   $J = (j_1, \dots , j_k)$ be a $k$-tuple of
increasing indices in $\{1,\dots,p\}$, with $k \leq p$. We define the
permutation importance of the sub-vector  $\mathbf X_{J} = (X_{j_1}, X_{j_2},\dots,X_{j_k})^\top$ of predictors by
\begin{equation*}
\mathcal I(\mathbf X_{J}) := \E \left[ \left( Y - f(\mathbf X_{(J)} ) \right)^2 \right] - \E \left[ \left( Y - f(\mathbf X )  \right)^2 \right],
\label{eq:grImp}
\end{equation*}
where $\mathbf X_{(J)}  = (X_1,\dots,X'_{j_1},X_{j_1 +1 }, \dots , X'_{j_2},X_{j_2 +1 }, \dots, X'_{j_\ell},X_{j_\ell +1 },\dots,X_p) ^\top $ is a random vector such that $\mathbf X'_J = (X'_{j_1}, X'_{j_2},\dots,X'_{j_k}) ^\top$ is an independent replicate of  $\mathbf X_{J}$, which is also independent of $Y$ and of all  other predictors. We call this quantity the grouped variable importance since it only depends on which variables appear in $\mathbf X_{J}$. By abuse of notation and ignoring the ranking, we may also refer to $\mathbf X_{J}$ as a group of variables.

\subsection{Decomposition of the grouped variable importance}

Let $\mathbf X_J$ be  a subgroup of variables from the random vector $\mathbf X$. Let $\mathbf X_{\bar J}$ denote the group of variables that does not appear  in $\mathbf X_J$. Assume that we observe $Y$ and $\mathbf X$ in the following additive regression model:
\begin{eqnarray}
Y &=& f  (\mathbf X) +  \varepsilon \notag \\
 &=&   f_{J} (\mathbf X_{J})  +  f_{\bar J} (\mathbf X_{\bar J})  + \varepsilon, \label{eq:addModel}
\end{eqnarray}
where the $f_{J}$ and $f_{\bar J}$ are two measurable functions and $\varepsilon$ is a random variable such that
$\E[\varepsilon|\mathbf X]=0$ and  $\E[\varepsilon^2|\mathbf X]$ is finite.
Results in \cite{rf:G+14} on  the permutation importance of individual variables can be extended to the case of a group of variables.
\begin{prop}
\label{prop:addModel}
Under model \eqref{eq:addModel}, the importance of  group $J$ satisfies
$$\mathcal I(\mathbf X_{J}) = 2 \Var \left[ f_{J} (\mathbf X_{J})  \right].$$
\end{prop}
The proof is given in~\ref{sub:proof}.
The next result gives the grouped variable importance for more specific models.
It can be easily deduced from Proposition~\ref{prop:addModel}.
\begin{cor} \label{cor:specificmodels}
Assume that we observe $Y$ and $\mathbf X$ in  model \eqref{eq:addModel} with
\begin{equation} \label{cond:addtit}
f_J (\mathbf x_J) = \sum_{j \in J} f_j (x_j),
\end{equation}
 where the $f_j$ are measurable functions and $\mathbf x_J = (x_j)_{j  \in J}$.
 \begin{enumerate}
 \item If the random variables $(X_j)_{j \in J}$ are independent, then
\begin{equation*}
\mathcal I(\mathbf X_{J}) = 2 \sum_{j \in J} \Var \left( f_j(X_j) \right)  =  \sum_{j \in J} \mathcal I \left( X_j
\right).
\end{equation*}
 \item If for any $j \in J$, $f_j$ is a linear function such that $f_j(x_j) =  \alpha_j x_j$, then
\begin{equation} \label{ImportanceCorrel}
\mathcal I(\mathbf X_{J}) = 2 \alpha_{J}^\top \Cov(\mathbf X_{J})  \alpha_{J},
\end{equation}
where $\alpha_{J} = (\alpha_{j})_{j \in J}$.
\end{enumerate}
\end{cor}
If  $f$ is additive and if the variables of the group are independent, the
grouped variable importance is nothing more than the sum of the individual
importances. As affirmed in the second point of Corollary~\ref{cor:specificmodels}, this property is lost as soon as the variables in the group are correlated.  The experiments presented in \ref{sub:ExpPIGV}  allow us to compare the grouped variable importance with the individual importances in various models. In essence, these experiments suggest that in general, the grouped variable importance is not comparable with the sum of the individual importances.  This is not surprising since the grouped variable importance is  a more accurate measure of the importance of a group of variables than a simple sum of the individual importances.

Next, let us introduce a rescaled version of the grouped variable importance:
$$ \mathcal I _{\textrm{nor}}(\mathbf X_{J}) := \dfrac{1}{|J|} \mathcal I(\mathbf X_{J}).$$
To see why this quantity makes sense, let us consider the situation where two groups of variables have almost the same
group permutation importances. Then, one would prefer to select first  the smaller group in order to obtain a sparser set of
predictors. We thus propose to normalise $ \mathcal I (\mathbf X_{J})$ by an increasing function of  group size $|J|$ so as to
compare permutation importances of groups of variables. Moreover, Corollary~\ref{cor:specificmodels} tells us
that  normalisation by $|J|$ is reasonable since it corresponds to comparing the means of the individual permutation
importances over each group in the case where the variables are independent.

In Section~\ref{sec:contrib}, we propose a variable selection algorithm based on the grouped variable importance. This algorithm uses the rescaled version to take into account group sizes in the selection process. More generally, we choose to work with the rescaled version when comparing groups of variables of different sizes.

\subsection{Grouped variable importance and random forests}
\label{sec:grRF}

Classification and regression trees are well-performing techniques for estimating $f$. A popular method in this context
is the CART algorithm of \citet{cart}. Though efficient,  tree methods  are also known to be unstable insofar as a
small perturbation of the training sample may change radically the predictions. To counter this,
\citet{rf:B01} introduced  random forests as a substantial improvement to decision trees.  The permutation
importance measure was also introduced in this seminal paper.
We now recall how individual  permutation importances can be estimated with random forests before presenting a natural extension to the estimation of grouped variable importances.

Assume that we observe $n$ i.i.d. replicates $\Dn = \ensemble{(\mathbf X_1, Y_1), \ldots, (\mathbf X_n, Y_n)}$ of $(\mathbf X, Y)$.  The random forest algorithm consists in aggregating a collection of random trees as in the bagging method, also proposed by \citet{bagging};  trees are built over $M$ bootstrap samples $\Dn^1, \ldots, \Dn^{M}$ of the training data $\Dn$. Instead of the CART algorithm, a subset of variables is randomly chosen for the splitting rule at each node. Each tree is then fully grown or grown until each node is pure, and not subsequently pruned. The resulting learning rule is the aggregation of all of the tree-based estimators, denoted by $\hat{f}_1, \ldots, \hat{f}_{M}$. In the regression setting,  aggregation is based on the average of the predictions.

For any $m \in \{1, \dots, M \}$, let $\Dnbar^m := \Dn \setminus \mathcal D_n^m$ be the corresponding out-of-bag sample.
The risk of $\hat f_m$ is estimated on the out-of-bag sample as follows:
$$\hat{R}(\hat f_m, \Dnbar^m) = \dfrac{1}{|\Dnbar^m|} \sum_{i: (\mathbf{X}_i, Y_i) \in \Dnbar^m} (Y_i - \hat  f_m(\mathbf{X}_i) )^2.$$
Let $ \Dnbar^{mj}$ be the permuted version of  $ \Dnbar^{m}$ obtained by randomly permuting the variable $X_j$ in each out-of-bag sample $ \Dnbar^{m}$. An estimation of the permutation importance measure of  variable $X_j$ is then given by
\begin{equation}
\hat{\mathcal I}(X_j) = \dfrac{1}{M} \sum_{m=1}^{M} \bigg[ \hat{R}(\hat{f}_m, \Dnbar^{mj}) - \hat{R} (\hat{f}_m, \Dnbar^{m}) \bigg].
\label{eq:empRawImp}
\end{equation}
This random permutation mimics the replacement of $X_j$ by $X_j'$ in \eqref{eq:rawImp} and breaks the link between $X_j$ and $Y$ and the other predictors.
We now extend the method for estimating the permutation importance of a group of variables $\mathbf{X}_J$. For any $m
\in \{1, \dots, M \}$, let $ \Dnbar^{mJ} $ be the permuted version of $ \Dnbar^{m}$ obtained by randomly permuting the
group $\mathbf{X}_J$ in each out-of-bag sample $ \Dnbar^{m}$. Note that the same random permutation is used for each
variable $X_j$ of the group. In this way the (empirical) joint distribution of $\mathbf{X}_J$ is left unchanged by the
permutation whereas the link between $\mathbf{X}_J$ and $Y$ and the other predictors is broken. The  importance of
$\mathbf{X}_J$ can be estimated by
\begin{equation}
\hat{\mathcal I}(\mathbf{X}_J) = \dfrac{1}{M} \sum_{m=1}^{M} \bigg[ \hat{R}(\hat{f}_m, \Dnbar^{mJ}) - \hat{R} (\hat{f}_m, \Dnbar^{m}) \bigg].
\label{eq:empGrImp}
\end{equation}
We define $\hat {\mathcal I} _{\textrm{nor}}(\mathbf X_{J})$ as the rescaled version of this estimate.
In the following section, we use the grouped variable importance as a criterion for selecting features in the context of multiple functional data.

\section{Multiple functional data analysis using  grouped variable importance}
\label{sec:contrib}

In this section, we consider an application of  grouped variable selection for multiple functional regression with scalar response $Y$. Each covariate $X^1, \dots, X^p$ takes its values in the Hilbert space $L^2([0,1])$ equipped with the inner product
$$\langle f,g \rangle_{L^2} = \int f(t) g(t) dt, $$
for $f, g \in L^2([0,1])$.
One common approach of functional data analysis is to project the variables onto a finite dimensional subspace of $L^2([0,1])$ and to use the basis coefficients in a learning algorithm \citep{fda:RS05}.
For instance, the wavelet transform is widely used  in signal processing and for nonparametric function estimation (see for instance
\citealp{wave:A+01}). Unlike Fourier bases and splines,  wavelets are localized both in frequency and time.

For $j\geq 0$ and $k = 0, \dots, 2^j-1$, define a sequence of functions $\phi_{jk}$ (resp. $\psi_{jk}$), obtained by translations and dilatations of a compactly supported function $\phi$ (resp. $\psi$), called a scaling function (resp. wavelet function). For any $j_0 \geq 0$, the collection
$$\Bcal = \{\phi_{j_0 k}, k = 0, \dots, 2^{j_0} - 1 \} \cup \{ \psi_{jk}, j \geq j_0, k = 0, \dots, 2^j-1 \} $$
forms an orthonormal basis of $L^2([0,1])$  (see for instance \citet{wave:PW00}). Then, a function $s \in L^2([0,1])$ can be decomposed as
\begin{equation}
s(t) = \sum_{k=0}^{2^{j_0} - 1} \langle s,\phi_{j_0 k} \rangle_{L^2} \phi_{j_0 k}(t) + \sum_{j\geq j_0} \sum_{k=0}^{2^j-1} \langle s, \psi_{jk} \rangle_{L^2} \psi_{jk}(t).
\label{eq:waveRepres}
\end{equation}
The first term in Equation~\eqref{eq:waveRepres} is the smooth approximation of $s$ at level $j_0$ while the second term is the detail part of the wavelet representation.
We assume that each covariate $X$ is observed  on a fine sampling grid $t_1, \dots, t_N$ with $t_\ell = \frac{\ell}{N}$. Note that a wavelet decomposition of $X$ can also be given in a similar form as in (\ref{eq:waveRepres}). For $j_0 =0$, we have
\begin{equation}
X(t_\ell) = \zeta \phi_{00}(t_\ell) + \sum_{j=0}^{J-1} \sum_{k=0}^{2^j-1} \xi_{jk} \psi_{jk}(t_\ell),
\label{eq:waveRepresTrunc}
\end{equation}
where $J:=\log_2(N)$ is the maximal number of wavelet levels and $\zeta$ and $\xi_{jk}$ are respectively the scale and
 wavelet coefficients of the discretized curve $X$ at  position $k$ for  resolution level $j$. These empirical
coefficients can be efficiently computed using the discrete wavelet transform  described in Chapter 4 of \citet{wave:PW00}.

\begin{figure}
	\begin{center}
	\includegraphics[width=0.4\textwidth]{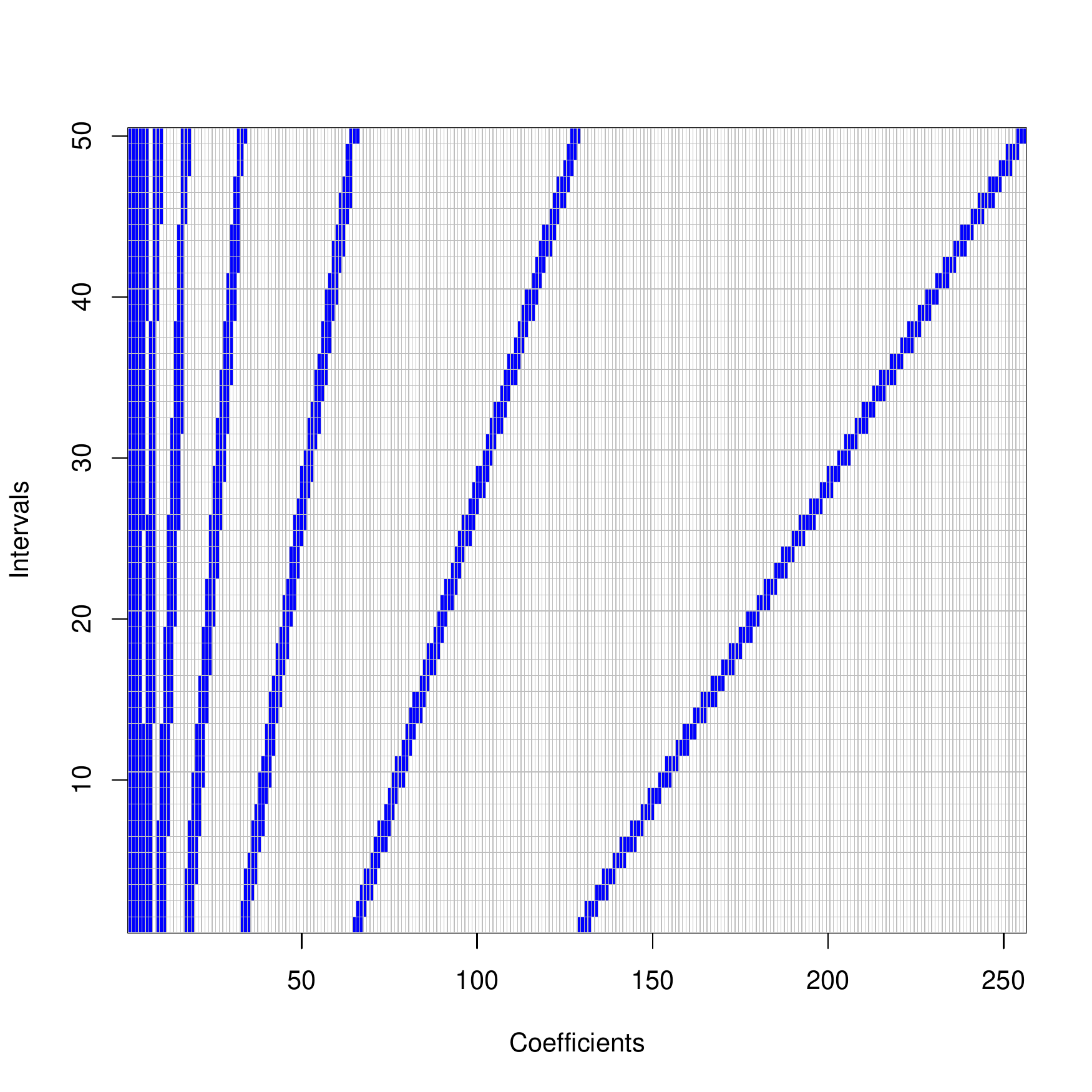}
	\caption{Correspondence between the time domain and  wavelet functions for a Daubechies wavelet basis with two vanishing moments. For the time $t$, the darkest points correspond to the wavelet functions which are non null at time $t$.}
	\label{fig:corresp}
	\end{center}
\end{figure}

For a given wavelet basis, we introduce the {\it wavelet support at time $t$} as the set of all  indices of wavelet
functions that are non null at $t$: $$\mathcal S(t) = \{(j, k): \psi_{jk}(t) \neq 0\}.$$  Figure~\ref{fig:corresp}
displays the matrix  giving the correspondence between a time  location and the associated wavelet functions for a
Daubechies wavelet basis with two vanishing moments.
In a similar way but for a time interval~$\mathcal T$, we define the
{\it wavelet support of  $\mathcal T$} by
\begin{eqnarray*}
\mathcal S(\mathcal T) &=&  \{(j, k): \psi_{jk}(t) \neq 0, \forall t\in \mathcal T \} \\
& = & \bigcap _{t \in  \mathcal T} \mathcal S(t) .
\end{eqnarray*}
This set corresponds to  wavelet functions localized on the interval $\mathcal T$.

\subsection{Grouped variable importance for functional variables}
\label{sec:varSelFunctData}

In this section, we show how the grouped variable importance can be fruitfully used for comparing the importance of different
wavelet coefficients in the context of functional prediction. Remember that $p$ functional covariates $X^1, \dots, X^p$
are observed, along with a scalar response $Y$. For the sake of simplicity,  covariates are decomposed on the same
wavelet basis $\Bcal$ but the methodology presented above could be also adapted with a specific basis for each
covariate.
For any $u \in \{1, \dots, p\}$, let $\mathbf W^u = (\zeta^u, \xi^u_{jk})_{jk}$ be the random vector composed of the wavelet coefficients of the functional variable $X^u$.

\subsubsection*{Groups of wavelet coefficients}

Wavelet coefficients are characterised by their frequency,  time location and  the functional variables they describe. Consequently, they can be grouped in many ways. We give below a nonexhaustive list of groups for which we are interested in computing the importance:
\begin{itemize}
\item {\bf A group related to a variable}. The vector $\mathbf W^u$ defines the  group $G(u)$.
\item {\bf A group related to a frequency level of a variable}. For a fixed variable $X^u $, the group is composed of
the wavelet coefficients of frequency level $j$:  $$G(j,u) := \{ \xi_{j,1}^u,\dots, \xi^u_{j,2^j-1} \} .$$
\item {\bf A group related to a frequency level}.  The group is composed of all the wavelet coefficients of frequency level $j$ for all the variables: $$G(j) :=   \bigcup _{u =1, \dots, p} G(j,u) .$$
\item {\bf A group related to a given time}. Define the group of ``active'' wavelet coefficients associated to a given time $t$ by $$ G(t) :=  \bigcup_{u = 1, \dots, p}  \{\zeta^u\}  \cup \bigcup_{u = 1, \dots, p, \, (j,k) \in \mathcal S(t)}    \{  \xi^u_{jk}  \}.$$ Depending on the size of the support of $\phi$ and $\psi$, the group $G(t)$ may be very large. For instance with a Daubechies wavelet basis with two vanishing moments, in Figure~\ref{fig:corresp} the group $G(t)$ is composed of the darkest points of the row corresponding to time $t$.
\item {\bf A group related to a time interval}. Let $[a,b]$ be a time interval. The group of ``active'' wavelet coefficients associated with $[a,b]$ is $$G([a,b]) := \bigcup _{t \in [a,b]} G(t) .$$
\end{itemize}
Many other groupings could be proposed. For instance, one could regroup pairs of correlated variables, or consider a group  composed of the wavelet coefficients taken in an interval of frequencies, a group related to a given time and a fixed variable, etc.

By computing the importance of such groups, one directly obtains a rough detection of the most important groups of coefficients for predicting $Y$.
When grouping by frequency levels or  time locations, all  groups do not have equal sizes. As explained in Section \ref{sec:RFgrimp}, it is preferable to use the rescaled version of the grouped variable importance in order to compensate the effect of group size in the grouped variable importance measure.

\subsubsection*{Grouped variable selection}

We now propose a more elaborate method for selecting groups of coefficients. The selection procedure is based on the Recursive Feature Elimination (RFE) algorithm proposed by \citet{svm:guyon2002} in the context of support vector machines. In this paper, we propose a random forests version of the RFE algorithm which is guided by  grouped variable importance. The procedure is summarized in Algorithm~\ref{algo:RFE}. This backward grouped elimination approach produces a collection of nested subsets of groups. The selected groups are obtained by minimizing the validation error computed in step~\ref{algo:error}.
\begin{algorithm}[h]
\caption{Grouped Variable Selection}
\begin{algorithmic}[1]
\STATE Train a random forest model \label{algo:train}
\STATE Compute the error using a validation sample \label{algo:error}
\STATE Compute the grouped variable importance measure
\STATE Eliminate the least important group of variables \label{algo:elim}
\STATE Repeat \ref{algo:train}--\ref{algo:elim} until no further groups remain
\end{algorithmic}
\label{algo:RFE}
\end{algorithm}

\noindent This  algorithm is motivated by the results of our previous work \citep{rf:G+14} about variable selection
using the permutation importance measure for random forests. Strong correlation between
predictors has a strong impact on the permutation importance measure.  It was also shown in this previous paper
that when  predictors are strongly correlated, the RFE algorithm provides a better performance than the
``non-recursive'' strategy (NRFE) that computes the grouped variable importance just once and does not recompute the
importance at each step of the algorithm. In the present paper,  we continue
this study by adapting the RFE algorithm for the grouped variable  importance measure.  We give below two
applications of this algorithm.
\begin{itemize}
\item {\bf Selection of functional variables}.  Each vector $\mathbf W^u$ defines  a group $G(u)$ and the goal is to perform a grouped variable selection over the groups $G(1), \dots, G(p)$. The selection  allows us to identify the most relevant functional variables.
\item {\bf Selection of  wavelet levels for a given functional variable}.   For a fixed $u$, we make a selection over the groups  $G(j,u)$ to identify the frequency levels which yield predictive information.
\end{itemize}
\begin{remark} \label{rem:time} Algorithm~\ref{algo:RFE} for grouped variable selection is appropriate for groups
defining a partition over the wavelet coefficients. This is not the case for groups related to time locations. The
algorithm can not easily be adapted to these groups  because most wavelet coefficients belong to several groups and
the elimination of a whole group may not be an efficient strategy. For instance the coefficient $\zeta^u$, which
approximates the smooth part of the curves and which is usually a good predictor, is common to all times $t$.
\end{remark}

 \subsection{Experiments}
\label{sec:simu}

In this section, we present various numerical experiments for illustrating  the interest of  grouped variable
importance for analyzing functional data. We first describe the simulation set-up.

\subsubsection*{Presentation of the general simulation design}

The experiments presented below consider one or several functional covariates for predicting an outcome  $Y$.
Except for the second simulation in Experiment 1 which is presented in detail in the next section, the functional
covariates  are  defined as functions of $Y$.

First, an $n$-sample of the outcome variable $Y$ is  simulated from a given distribution specified for each experiment.
The realization of a functional covariate $X$ (denoted by $X^u$ when there are several) is an
$n$-sample of independent discrete time random processes $X_i = ( X_i(t_\ell) )_{\ell = 1, \dots, N}$, for all  $i \in
\{1,\dots,n\}$ under a model of the form
\begin{equation}
X_{i}(t_\ell) = s(t_\ell,Z_i) + \sigma \varepsilon_{i,\ell}, \; \ell=1,\dots,N,
\label{eq:simu}
\end{equation}
where the $\varepsilon_{i,\ell}$ are i.i.d standard Gaussian random variables, and $t_\ell = \frac \ell N$. The random
variable $Z_i$ is correlated with $Y_i$ and will be specified for each experiment; it is equal to the outcome variable
$Y_i$  for most of the experiments.  The functional covariates are actually simulated in the wavelet domain from the
following model:
for $i=1,\dots,n$,  $j = 0, \dots, J-1$ and $k = 0, \dots, 2^j-1$,
\begin{equation}
\zeta_i = \omega_0 + h_{\zeta}(Z_i) + \sigma \eta_{i\zeta} ,
\label{eq:simScaling}
\end{equation}
and
\begin{equation}
\xi_{ijk} =
\left\lbrace
\begin{array}{llll}
\omega_{jk} + h_{jk}(Z_i) + \sigma \eta_{ijk}   &  \mbox{ if } j \leq j^\star, \; k \in \{ 0, \dots, 2^j-1\},\\
0  &  \mbox{ if }     j^\star < j \leq J -1,
\end{array}
\right.
\label{eq:simCoeff}
\end{equation}
where  $j^\star$ is the highest wavelet  level of the signal.
The random variables $\eta_{ijk}$ and $\eta_{i\zeta}$ are
i.i.d. standard Gaussian. The ``signal'' part of
Equation~\eqref{eq:simCoeff} is the sum of a random coefficient $\omega_{jk}$ whose realisation is the same for all $i$,
and a link function $h_{jk}$. The  coefficients $\omega_0$ and $\omega_{jk}$ in  \eqref{eq:simScaling}   and
\eqref{eq:simCoeff} are simulated as follows:
\begin{equation*}
\left\lbrace
\begin{array}{llll}
\omega_{jk} &\sim& \mathcal N_1(0, \tau_j^2), & \mbox{if }j \leq j^\star, \; k \in \{ 0, \dots, 2^j-1\},\\
\omega_0 &\sim& \mathcal N_1(3,1),
\end{array}
\right.
\end{equation*}
where $\tau_j = e^{-(j-1)}$. Note that the standard deviation $\tau_j$
decreases with $j$ and thus less noise is added to the first wavelet levels. The link function $h_{jk}$ describes the
link between the wavelet coefficients $\xi_{ijk}$ and the variable $Z$ (or  the outcome variable $Y$). Two
different link functions are considered in the experiments:
\begin{itemize}
\item a linear link: $h_{jk}(z) = \theta_{jk} z$,
\item  a logistic link: $h_{jk}(z) = \dfrac{\theta_{jk}}{1+e^{-z}}$,
\end{itemize}
where the coefficients $\theta_{jk}$  parametrize  the strength of the relationship between $Z$ and the
wavelet coefficients. The $n$ discrete processes $X_1, \dots, X_n $ are simulated according to
 \eqref{eq:simScaling} and
\eqref{eq:simCoeff} before applying the inverse wavelet transform.
We choose a Daubechies wavelet filter with four vanishing moments to simulate the observations. We use the same
basis for the projection of the functional observations.

\subsubsection*{Experiment 1: Detection of important time intervals}

In this first experiment, we illustrate the use of the grouped variable importance for  detection of the most
relevant time intervals.
We simply estimate the importance of time intervals without applying Algorithm~\ref{algo:RFE} (see
Remark~\ref{rem:time}). We only consider one functional covariate $X$ since it will be sufficient to illustrate the
method. Let  $\mathcal T^\star = [t_{50}, t_{55}]$, we propose two simulation designs for which the outcome $Y$ is
correlated to the signal $X$ on the interval $\mathcal T^\star$.
\begin{itemize}
\item {\bf Simulation 1}.  For this first simulation, we follow the general simulation design presented before by
considering linear link functions $h$ for all  wavelet coefficients belonging to the wavelet support $\mathcal S
(\mathcal T^\star)$. The outcome variable $Y$ is simulated from a Gaussian distribution $\mathcal{N}_1(0, 3)$.
We simulate the wavelet coefficients as in \eqref{eq:simCoeff} and the scaling coefficients as in
\eqref{eq:simScaling} with $Z=Y$. We take linear link functions and set $n=1000$, $\sigma = 0.01$, $N= 2^8
$ and thus
$J=8$. We take $j^\star =7$ which means that even the wavelet coefficients of highest level $j = 7$  are not Dirac
distributions at zero. The wavelet coefficients and the scaling coefficient
are generated as follows: for any $j \in \{0, \dots, J-1\}$ and any $k \in  \{0, \dots, 2^j-1\}$,
\begin{equation}
\xi_{ijk} = \left\lbrace
\begin{array}{lll}
\omega_{jk} +   Y_i + \sigma \eta_{ijk}, & \mbox{if } (j,k) \in \mathcal S(\mathcal T^\star)\\
\omega_{jk} + \sigma \eta_{ijk}, & \mbox{otherwise,}
\end{array}
\right.
\label{eq:ex3Wave}
\end{equation}
and
\begin{equation}
\zeta_i = \omega_0 + Y_i + \sigma \eta_{i\zeta},
\label{eq:ex3Scaling}
\end{equation}
for $i=1,\dots,n$.
\item {\bf Simulation 2}. In contrast to the previous simulation, we first simulate the functional variable $X$ and then
simulate the outcome variable $Y$ as a function of $X$. The functional variable is simulated in the wavelet domain
according to   \eqref{eq:simScaling} and \eqref{eq:simCoeff} with $h_{jk} = h_{\zeta}= 0$ for all $j$ and $k$. We
also take $n=1000$, $\sigma = 0.01$, $N= 2^8 $ and $j^\star =7$. By applying the wavelet inverse transform for any $i$,
we obtain an $n$-sample of discrete time random processes $X_i = ( X_i(t_\ell) )_{\ell = 1, \dots, N}$. Figure~\ref{fig:exSignals} displays a set of ten of these processes.

\begin{figure}[h]
	\begin{center}
	\includegraphics[width=0.4\textwidth, page=1]{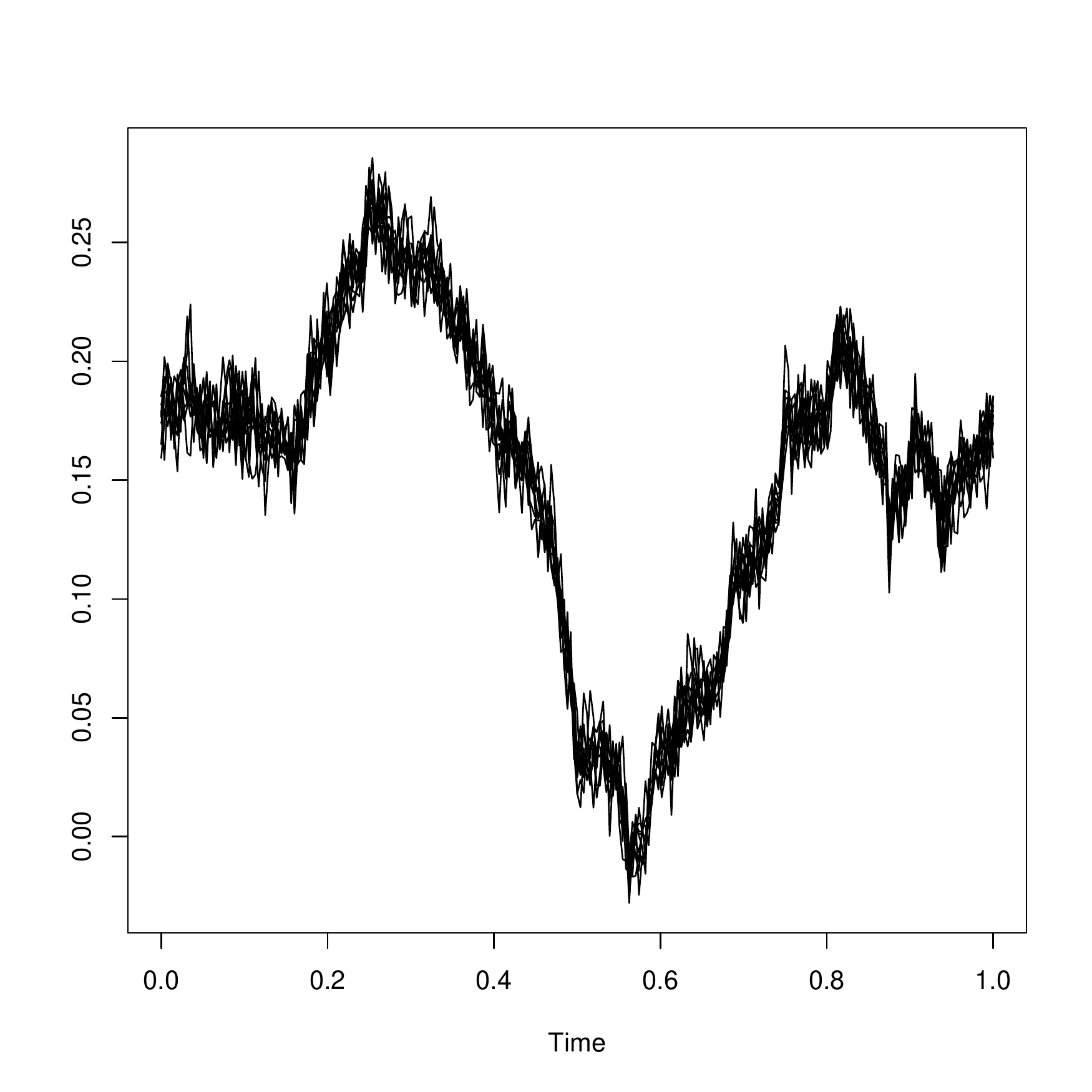}
	\caption{Experiment 1 -- Example of ten processes drawn from the protocol used for Simulation 2.}
	\label{fig:exSignals}
	\end{center}
\end{figure}

The outcome variable  $Y$ is then obtained via
$$Y_i = \dfrac{1000}{|\mathcal T^\star|} \sum_{t_\ell \in \mathcal T^\star} |X_i(t_\ell) - X_i(t_{\ell-1})|.$$
Thus, $Y_i$ is a measure of the oscillations of the curve $X_i$ over the interval $\mathcal T^\star$.
\end{itemize}
The aim is to detect  $\mathcal T^\star$ using the grouped variable importance.  In both cases, the grouped variable
importance $\mathcal I (G(t))$ is evaluated at 50 equally spaced time points. Figure~\ref{fig:impTemp} displays the
importance of the time points, averaged over 100 iterations. The first and third
quartiles are also represented in order to highlight  estimation variability. In the two cases, the importance estimation
makes it possible to detect $\mathcal T^\star$.

Note that the detection problem is tricky in the second case because the link between $Y$ and the wavelet coefficients
is complex here. Consequently the estimated importances are low
and the important intervals are difficult to detect.
\begin{figure}[h]
	\begin{center}
	\subfloat[Case
1]{\includegraphics[width=0.4\textwidth, page=1]{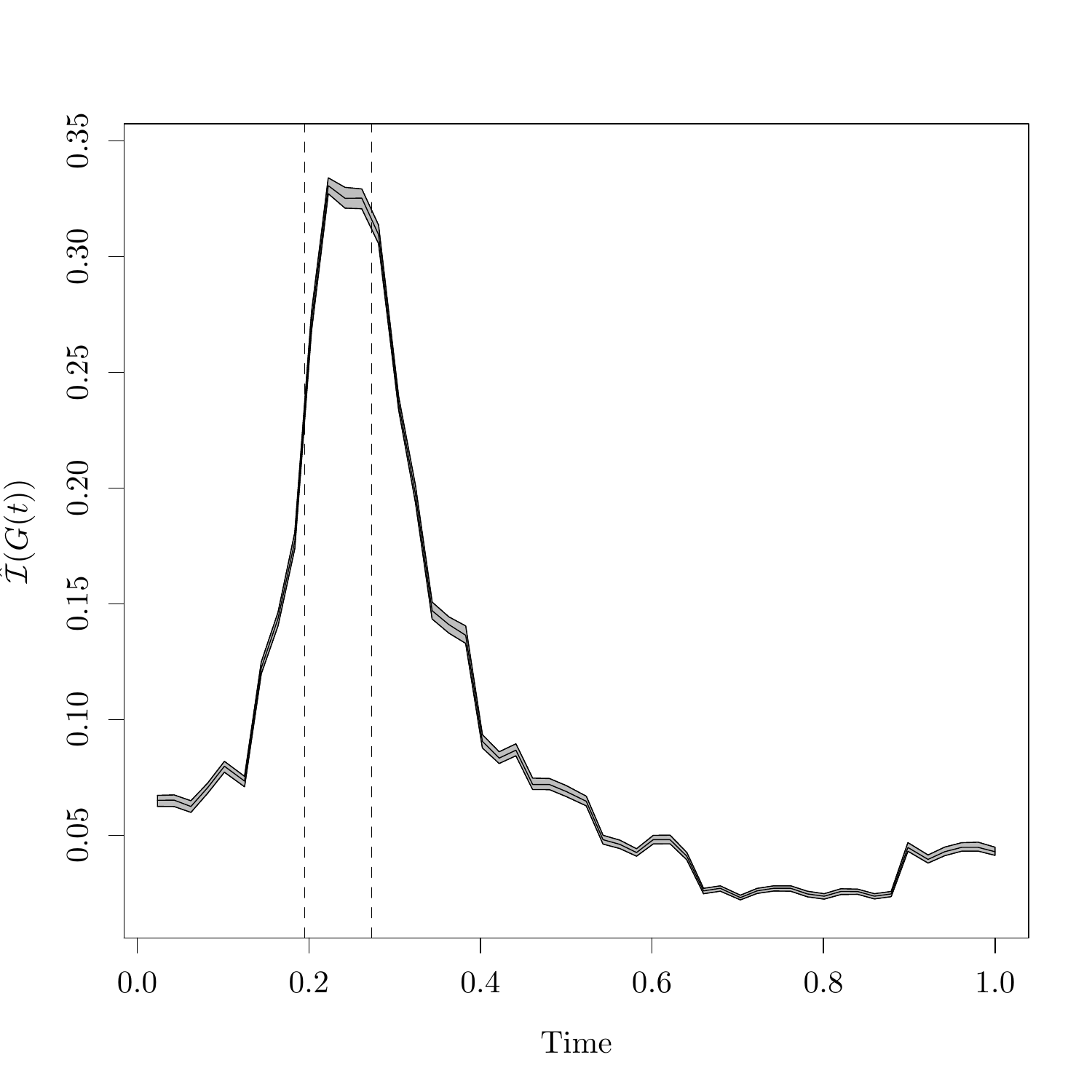}\label{fig:impTempCase1}}
	\subfloat[Case
2]{\includegraphics[width=0.4\textwidth, page=2]{impTemp.pdf}\label{fig:impTempCase2}}
	\caption{Experiment 1 -- Averaged time importances and first and third quartiles over 100 iterations. The  time
interval  $\mathcal T ^\star$ is located between the two vertical lines.}
	\label{fig:impTemp}
	\end{center}
\end{figure}

\subsubsection*{Experiment 2: Selection of wavelet levels}

This simulation deals with the selection of wavelet levels for one functional variable. We follow the general simulation
design presented above. The outcome variable $Y$ is simulated from a Gaussian distribution $\mathcal{N}_1(0, 3)$. We
simulate the wavelet coefficients as in \eqref{eq:simCoeff} and the scaling coefficients as in
\eqref{eq:simScaling} with $Z=Y$. We perform two types of simulation:
\begin{itemize}
\item In the first case we use linear link functions:
\begin{equation*}
h_{\zeta}(y) = 0.1 y
\end{equation*}
and
\begin{equation*}
h_{jk}(y) = \left\lbrace
\begin{array}{lll}
\theta_j y & \mbox{ if } j \leq 3, \; k  \in \{ 0, \dots, 2^j-1\}, \\
0 	 & \mbox{ otherwise,}
\end{array}
\right.
\end{equation*}
where the $\theta_j$ decrease linearly  from 0.1 to 0.01.
\item For the second simulation, we use logistic link functions:
\begin{equation*}
h_{\zeta}(y) =\frac{ 0.1}{1+e^{-y}}
\end{equation*}
and
\begin{equation*}
h_{jk}(y) = \left\lbrace
\begin{array}{lll}
\dfrac{\theta_{jk}}{1+e^{-y}}  & \mbox{if } j \leq 3, \;  k \in \{ 0, \dots, 2^j-1\}, \\
0 	 & \mbox{ otherwise,}
\end{array}
\right.
\end{equation*}
where the $\theta_j$ decrease linearly  from 0.1 to 0.01.
\end{itemize}
We fix $n=1000$, $\sigma = 0.05$, $N= 2^8 $ (thus $J=8$) and  $j^\star =7$ in both  cases.
The aim is to identify the most relevant wavelet levels for the prediction of $Y$ using the grouped importance. We
regroup the wavelet coefficients by wavelet levels: for $j \in \{0,\dots,J-1\}$,
$$ G(j) = \left\{ \xi_{jk}, k \in \{1,\dots,2^j-1\} \right\} $$
and
$$G_\zeta = \{ \zeta \}.$$
We apply  Algorithm~\ref{algo:RFE} with these groups. As the group sizes are  different, the
rescaled grouped importance criterion given in Section~\ref{sec:grRF} is used.

\begin{figure}[h]
	\begin{center}
	\subfloat[]{\includegraphics[width=.33\textwidth,
page=1]{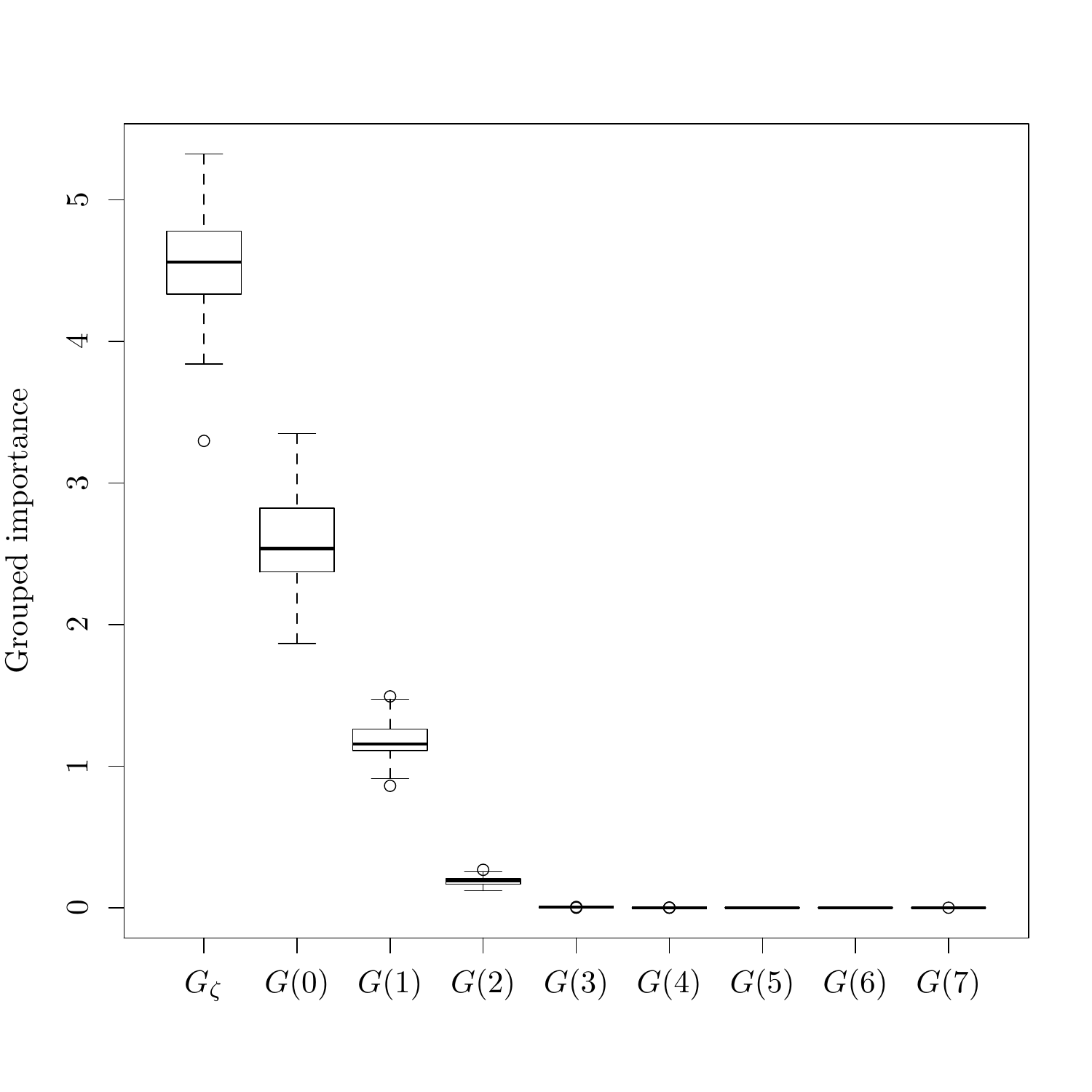}\label{fig:levBoxplots}}
	\subfloat[]{\includegraphics[width=.33\textwidth,
page=2]{level_tikz.pdf}\label{fig:levError}}
	\subfloat[]{\includegraphics[width=.33\textwidth,
page=3]{level_tikz.pdf}\label{fig:levFreq}}
	\caption{Experiment 2, linear links -- selection of the wavelet levels. From  left to  right: (a)
Boxplots of the grouped variable importances, (b) MSE error versus the number of groups and (c) Selection
frequencies.}
	\label{fig:lev}
	\end{center}
\end{figure}

\begin{figure}[h]
	\begin{center}
	\subfloat[]{\includegraphics[width=.33\textwidth,
page=1]{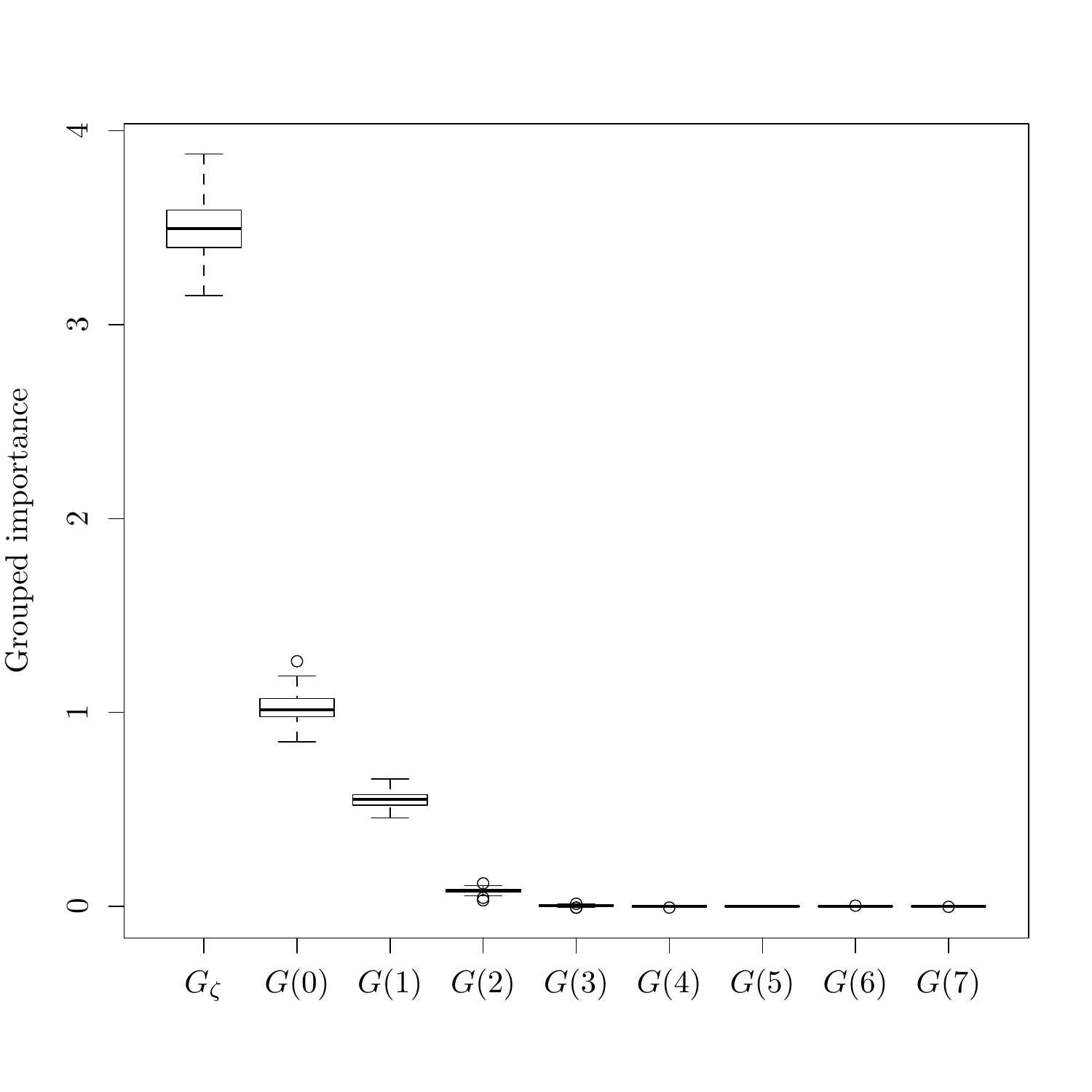}\label{fig:levBoxplotsLogit}}
	\subfloat[]{\includegraphics[width=.33\textwidth,
page=2]{levelLogit_tikz.pdf}\label{fig:levErrorLogit}}
	\subfloat[]{\includegraphics[width=.33\textwidth,
page=3]{levelLogit_tikz.pdf}\label{fig:levFreqLogit}}
	\caption{Experiment 2, logistic links -- selection of the wavelet levels.  From  left to  right: (a)
Boxplots of the grouped variable importances, (b) MSE error versus the number of groups and (c) Selection
frequencies.}
	\label{fig:levLogit}
	\end{center}
\end{figure}

The  trials are both repeated 100 times.  Figures~\ref{fig:lev} and \ref{fig:levLogit} respectively  give the
results for the linear  and  logistic links.
Let us first look at the trials with a linear link.  The boxplots of the grouped permutation importances at the first step of
the algorithm over the 100 trials are given in Figure~ \ref{fig:levBoxplots}. The fifth group  $G(3)$ being  not
strongly correlated with $Y$, its importance is  close to zero. It is selected  40 times out of the 100 simulations
(Fig.~\ref{fig:levFreq}) whereas $G_\zeta$, $G(0)$, $G(1)$ and $G(2)$ are almost always selected.
The other groups are not correlated with $Y$ and are almost never selected. For each trial, the mean squared
error (MSE) is computed as a function of the number of variables in the model. Figure~\ref{fig:levError} shows the average
of the errors over the 100 simulations.  On average, the model selected by minimizing the MSE  includes
four groups  but the model with five groups also has an error close to the minimum.

The simulation with the logistic link gives similar results. However the fifth
group $G(3)$  is more frequently selected (Fig.~\ref{fig:levFreqLogit}). The minimization
of the MSE leads to the selection of  five groups as shown in Figure~\ref{fig:levErrorLogit}. Note that this
random forest and grouped variable importance-based approach performs well even with a nonlinear link.

In both experiments,  the grouped variable importances obtained at the first step of the algorithm are ranked in the
same order as the $\theta_j$. Indeed the impact of the correlation between predictors is almost null in both
cases because of the orthogonality of the wavelet bases. In this context, the backward Algorithm~\ref{algo:RFE} does
not provide additional information compared to the
``non-recursive'' strategy (see the discussion following Algorithm~\ref{algo:RFE}).

\subsubsection*{Experiment 3: Selection of functional variables in the presence of strong correlation}

This simulation  illustrates the interest of Algorithm~\ref{algo:RFE} when selecting functional variables in the presence
of correlation.
First,  we simulate  $n=1000$ i.i.d. realizations of $p=10$ functional variables $X^1, \dots, X^{p}$  according to the
general simulation design detailed before. For all $i \in  \{1,\dots,n\}$, let $Z_i^1, \dots, Z_i^p$ be  latent
variables drawn from a standard Gaussian distribution.
The outcome variable $Y$ is defined as:
$$Y_i =  3.5 \, Z^1_i + 3 \, Z^2_i + 2.5 \, Z^3_i + 2.5 \, Z^4_i.$$
Then for $u \in \{1,\dots,p\}$, the wavelet coefficients are simulated according to \eqref{eq:simCoeff} and
\eqref{eq:simScaling} with a linear link:
\begin{equation*}
\xi^u_{ijk} =
\omega^u_{jk} + Z_i^u + \sigma \eta^u_{ijk} \quad  \mbox{ if } j \leq  j^\star,  \; k  \in \{0, \dots, 2^j-1  \}
\end{equation*}
and
$$\zeta^u_i = \omega^u_0 +  Z_i^u + \sigma \eta^u_{i\zeta},$$
with $\sigma = 0.1$, $N= 2^9$ and thus $J= 9$. We take $j^\star=3$ in order to make the functional variables smooth
enough.  Among the 10 variables $X^u$, the first four variables have decreasing predictive power whereas the others
are independent of $Y$.  Next, we add $q=10$ i.i.d. variables $X^{1,1}, \dots, X^{1,q}$ which are strongly correlated
with $X^1$:
for any $v \in  \{1, \dots, q\}$, any $i \in  \{1, \dots, n\}$,
$$\zeta^{1,v}_i =  \zeta^1_i  +  \tilde \sigma \eta^{1,v}_{i\zeta}$$
and
$$\xi_{ijk}^{1,v} =
\left\lbrace
\begin{array}{lll}
\xi_{ijk}^1 + \tilde \sigma \eta^{1,v}_{ijk} & \mbox{if }   j \leq j^\star, \;   k \in \{0, \dots, 2^j-1\}   \\
0 & \mbox{otherwise,}
\end{array}
\right.
 $$
where $\tilde \sigma=0.05$. The $\eta^{1,v}_{ijk}$  and the $\eta^{1,v}_{i\zeta}$  are i.i.d. standard Gaussian
random variables. The discrete processes $X_i^{1,v}$ are obtained using the inverse wavelet transform.
In the same way, we add $q=10$ i.i.d. variables $X^{2,1}, \dots, X^{2,q}$ which are strongly correlated with $X^2$. To sum up, the vector of predictors is composed of 30 variables:
$$X^1,X^{1,1}, \dots, X^{1,q},X^2, X^{2,1}, \dots,
X^{2,q},X^3, X^4,\dots, X^p. $$

\begin{figure}[h]
	\begin{center}
	\subfloat[]{\includegraphics[width=.33\textwidth,
page=1]{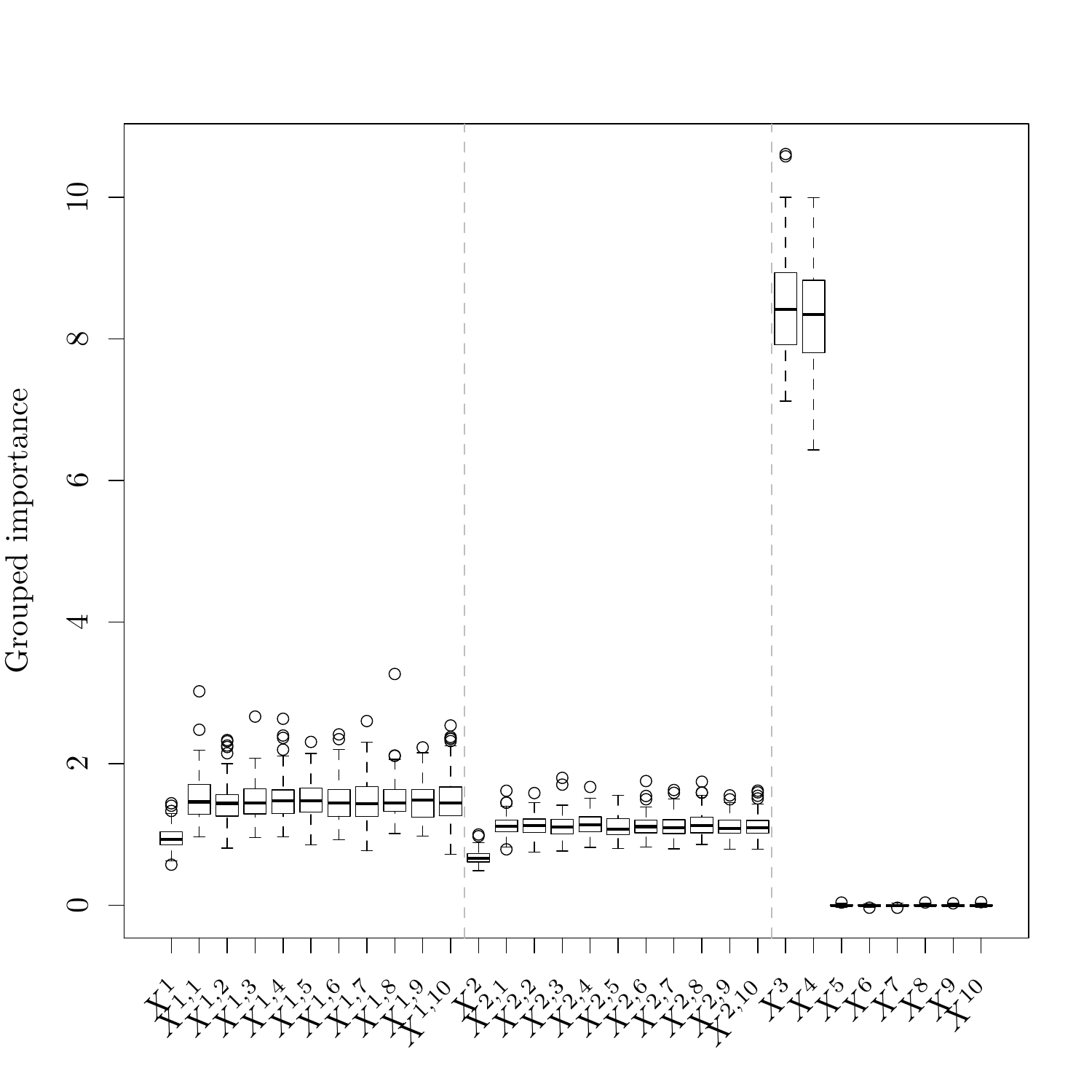}\label{fig:variableCorrelatedBP}}
	\subfloat[]{\includegraphics[width=.33\textwidth,
page=2]{variable_correlated_tikz_run10.pdf}\label{fig:variableCorrelatedErr}}
	\subfloat[]{\includegraphics[width=.33\textwidth,
page=4]{variable_correlated_tikz_run10.pdf}\label{fig:variableCorrelatedFreq}}
	\caption{Experiment 3 -- Selection of  functional variables.  From  left to right: (a)
Boxplots of the grouped variable importances, (b) MSE error versus the number of groups and (c) Selection
frequencies using Algorithm~\ref{algo:RFE}.}
	\label{fig:variableCorrelated}
	\end{center}
\end{figure}

The aim is to identify the most relevant functional variables for the prediction of $Y$. For each functional variable,
we regroup all  wavelet coefficients  and  apply Algorithm~\ref{algo:RFE}. This process is repeated 100 times.
Boxplots of the group permutation importances at the first step of the algorithm, over the 100 trials and for
each functional variable, are given in Figure~\ref{fig:variableCorrelatedBP}.
We see that the importances of the variables $X^1$ and $X^2$ and their noisy replicates are much lower than those
of  variables $X^3$ and $X^4$. This is due to the strong correlations between the two first variables and their
noisy replicates.
Indeed, the importance measure decreases when the correlation or the number of correlated variables increase. Note that the importances of $X^1$ and $X^2$ are slightly lower than those of their noisy replicates. This can be explained by the fact that the correlation between $X^1$ and its noisy replicates is higher than, for instance, the correlation of $X^{1,1}$ with $X^1, X^{1,2}, \dots, X^{1,q}$.

Figure~\ref{fig:variableCorrelatedErr} gives a comparison of the performance of Algorithm~\ref{algo:RFE} with the
``non-recursive'' strategy (NRFE).  Algorithm~\ref{algo:RFE} clearly shows better prediction performances. In
particular, it reaches a minimum error faster than the NRFE: only five variables for Algorithm~\ref{algo:RFE} whereas NRFE needs about twelve.
The RFE procedure is more efficient than the NRFE when  predictors are highly correlated.

Additional information is displayed in Figure~\ref{fig:variableCorrelatedFreq}. The selection frequencies using
Algorithm~\ref{algo:RFE} show that
the variables $X^3$ and $X^4$ are always selected. Indeed, these two variables have predictive power and they are not
correlated with the other predictors.  Note that $X^1$ and $X^2$ are  selected less often than their replicates,
even though they are more correlated with $Y$. This also comes from the fact that the
correlation between $X^1$ and its replicates is higher
than the correlation of $X^{1,1}$ with $X^1, X^{1,2}, \dots, X^{1,q}$.
We observe that $X^1$ and $X^2$ are eliminated in the first steps of the backward procedure, but this has no
consequence on
the prediction performance of  Algorithm~\ref{algo:RFE}.
These results motivate the use of this algorithm in practice: it reduces the effect of the correlation between
predictors on the selection process and  provides better prediction performances.

\section{A case study: variable selection for aviation safety}
\label{sec:appli}

In this section, we study a real problem coming from aviation safety. Airlines collect large amounts of information during
flights using flight data recorders. For several years now, airlines are required to use these data for flight safety purposes. A
large number of flight parameters (up to 1000) are recorded each second, including  aircraft speed,
accelerations, the heading, position, and warning signals. Each flight provides a multivariate
time series corresponding to this family of functional variables.

We focus here on the risk of long landing. A  data sequence for $N=512$ seconds before touchdown is observed for predicting the
landing distance. The evaluation of the risk of long landings is crucial for safety managers to avoid runway excursions
and more generally  to keep a high level of safety.  One answer to this problem is to select the flight parameters that
best explain  the risk  of long landings. Therefore, we attempt to find a sparse model showing good predictive
performances. In the future, flight data analysis could be used for pilot training or development of new
flight procedures during approach.

Following  aviation experts, 23 variables are preselected, see Table~\ref{tab:flightparam}. A sample of 1868 flights from the same airport and the
same company is considered. The functional variables are projected on a Daubechies wavelet basis with four vanishing
moments using the discrete wavelet transform algorithm as in Section~\ref{sec:contrib}. The choice of the wavelet basis
is determined by the nature of the flight data, which here contains information on both the time and
frequency scales.

\begin{table}[htdp] 
\begin{center}
\begin{tabular}{|c|c|}
\hline
Abbreviation  & Flight parameter \\ \hline
 AOA &  angle of attack  \\
 ALT.STDC &  altitude \\
 CASC &  airspeed \\
 CASC.GSC &  wind speed \\
 DRIFT &  aircraft drift angle \\
 ELEVR, ELEVL &  elevator position \\
 FLAPC &  flaps position \\
 GLIDE.DEVC &  deviation over the glide path \\
 GW.KG &  gross weight \\
 HEAD.MAG &  magnetic heading \\
 IVV &  inertial vertical speed \\
 N11C, N12C, N21C, N22C &  engine rotation speed \\
 PITCH &  rotation on the lateral axis \\
 PITCH.RATE &  aircraft pitch rate \\
 ROLL &  rotation on the longitudinal axis \\
 RUDD &  rudder position \\
 SAT &  static air temperature \\
 VAPP &  approach speed \\
 VRTG &  vertical acceleration \\
 \hline
\end{tabular} 
\end{center} 
\caption{List of preselected flight parameters.} \label{tab:flightparam}
\end{table}

\subsubsection*{Preliminary dimension reduction}

The design matrix formed by the wavelet coefficients for all of the chosen flight parameters has dimension $23 \times 512 =
11~776$. Selecting the variables directly from the whole set of coefficients is computationally prohibitive, so we first need to
significantly reduce the dimension. The naive method that shrinks the $n$ curves independently according to \citet{wave:DJ94}
and then brings the non-zero coefficients together in a second step would lead us to consider a large block of
coefficients with many zero values. This  is not relevant in our context, so we  propose an alternative
method which consists of shrinking the wavelet coefficients of the $n$ curves simultaneously.  More
precisely,  this method is  adapted  from \citet{wave:DJ94} for the particular context of $n$
independent (but not necessary identically distributed) discrete random processes. Shrinkage is performed on the norm
of the $n$-dimensional vector containing the wavelet coefficients. The complete method is described in
\ref{sec:dimRed}.

\subsubsection*{Selection of flight parameters}
\begin{figure}
	\begin{center}
	\subfloat[]{\includegraphics[width=.4\textwidth]{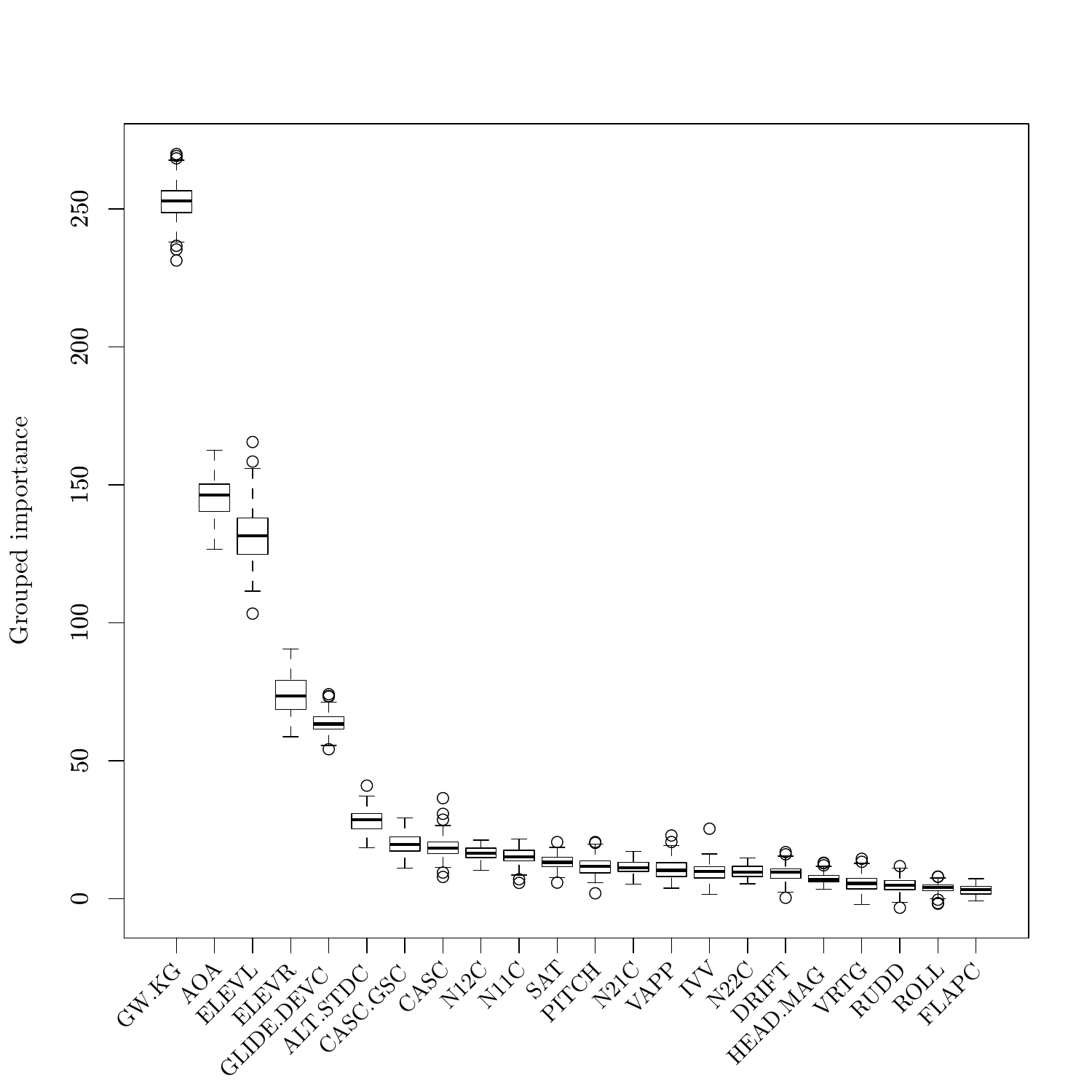}\label{fig:longLandingBoxplot}}
	\subfloat[]{\includegraphics[width=.4\textwidth, page=2]{longLanding_tikz.pdf}\label{fig:longLandingMSE}}
	\hfill
	\subfloat[]{\includegraphics[width=.4\textwidth, page=3]{longLanding_tikz.pdf}\label{fig:longLandingfreq}}
	\caption{Application to  long landings -- from  left to  right: (a) Boxplots of the grouped variable
importance, (b) MSE error versus the number of groups and (c) selection frequencies.}
	 \label{fig:longLanding}
	\end{center}	
\end{figure}

We obtain a selection of the functional parameters by  grouping  together the wavelet coefficients of each  flight
parameter and applying Algorithm~\ref{algo:RFE} with these groups. At each iteration, we randomly split the dataset into
a training set containing 90~\% of the data and a validation set containing the remaining 10~\%. In the backward
algorithm, the grouped variable importance is computed on the training set and the validation set is only used to compute
the MSE errors. The selection procedure is repeated 100 times to reduce the variability of the selection.
The final model is chosen by minimizing the averaged prediction error. Figure~\ref{fig:longLandingBoxplot} represents the boxplots of the grouped variable importance values computed on the 100 runs of the selection algorithm. According to this ranking, five variables are found to be significantly relevant. Looking at the averaged MSE estimate on Figure~\ref{fig:longLandingMSE}, we see that the average number of selected variables is ten, but  taking only five   is sufficient to get  a risk close to the minimum.

Figure~\ref{fig:longLandingfreq} gives additional information, displaying the proportion of times each flight parameter is selected. First, it confirms the previous remarks: five variables are always selected by the algorithm and the first ten  are selected more than 60 times over the 100 runs.
Second, it shows that  flight parameters related to  aircraft trajectory during  approach are among the most relevant  for predicting  long landings.
Indeed, the elevators (ELEVL, ELEVR) are used by the pilots to control the pitch of the aircraft. These have an effect on the angle of attack (AOA) and consequently on the landing. The variable GLIDE.DEVC is the glide slope deviation, i.e., the deviation between the aircraft trajectory and a glide path of approximatively three degrees above horizontal.
Other significant variables are the gross weight (GW.KG) which has an effect on the deceleration efficiency, the airspeed (CASC) and the engine rating (N11C, N12C).

\begin{figure}
	\begin{center}
	\subfloat[]{\includegraphics[width=.4\textwidth]{longLanding_grImp.pdf}\label{fig:longLandingBoxplot2}}
	\subfloat[]{\includegraphics[width=.4\textwidth]{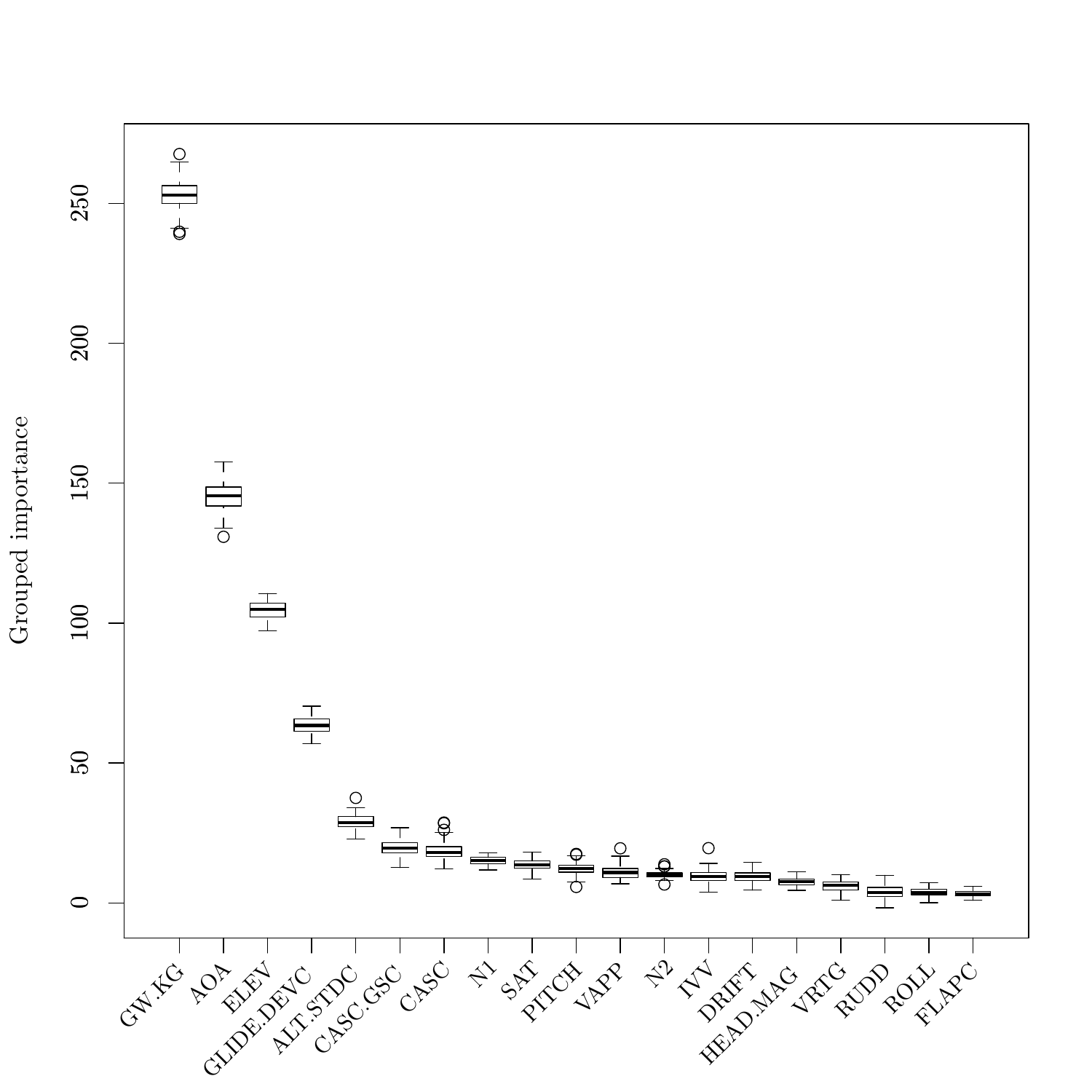}}
	\caption{Application to  long landings -- grouped variable importance measured before and after grouping the
correlated flight parameters N11C and N12C, N21C and N22C, ELEVR and ELEVL, as N1, N2 and ELEV.}
	 \label{fig:compareCorrelationFdata}
	\end{center}	
\end{figure}

It should be noted that the ranking based on selection frequency is close to the direct ranking given by the
importance measures when all  variables are included in the model. This suggests that  for this data,  correlation
between  predictors  is not strong enough to influence their  permutation importance measures. Moreover, if we
regroup several variables such as for example  flight parameters N11C and N12C, N21C and N22C, and ELEVR and ELEVL into
three new variables N1, N2 and ELEV, Figure~\ref{fig:compareCorrelationFdata} shows that the ranking remains unchanged.

\subsubsection*{Selection of wavelet levels}
\begin{figure}
	\begin{center}
	\subfloat[Gross weight -- MSE versus the number of groups]{\includegraphics[width=.4\textwidth, page=1]{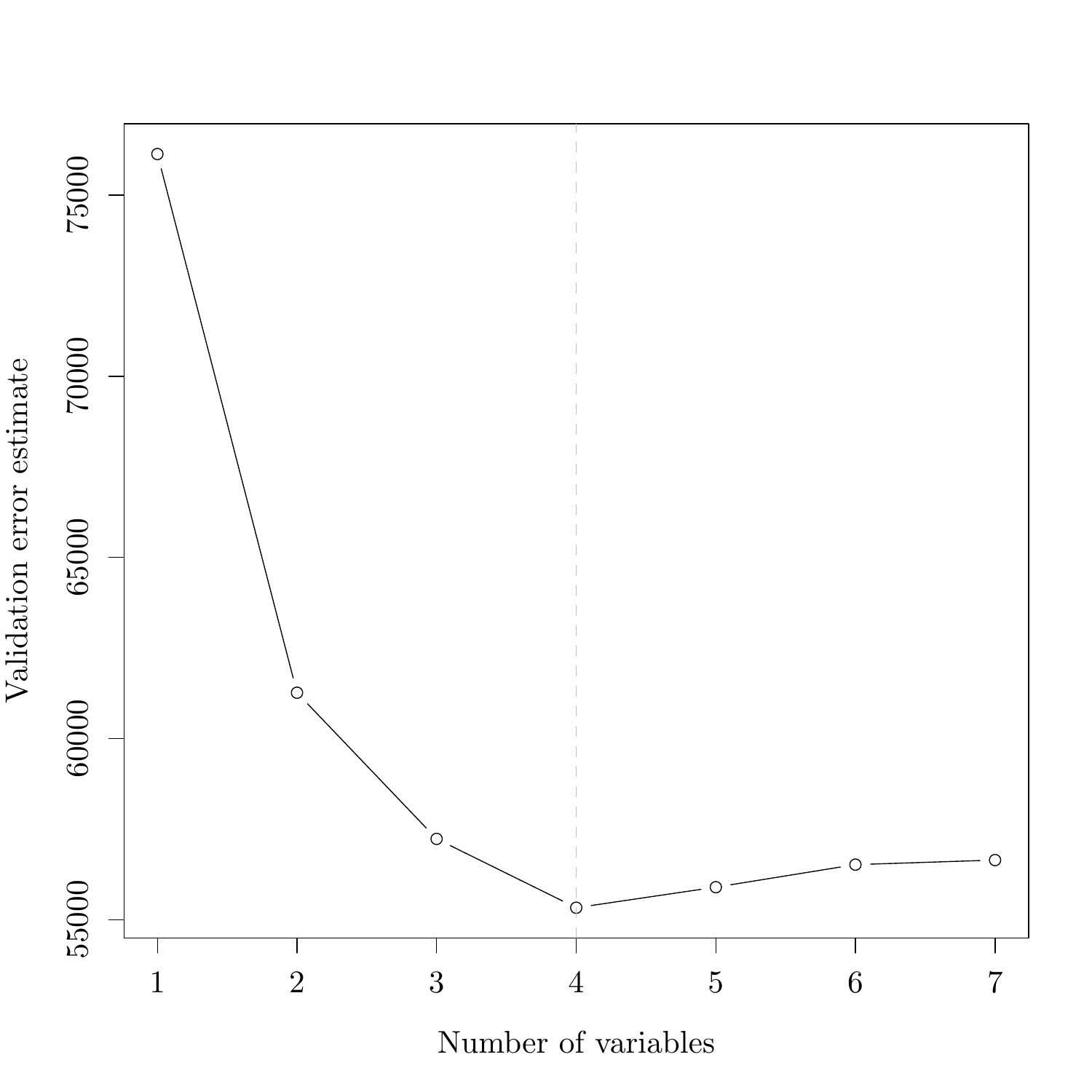}\label{fig:longLandingMSE_levGW}}
	\subfloat[Gross weight -- Selection frequencies]{\includegraphics[width=.4\textwidth, page=2]{levSel_longLandingEAP_GW_KG_tikz.pdf}\label{fig:longLandingfreq_levGW}}
	\hfill
	\subfloat[Angle of attack -- MSE versus the number of groups]{\includegraphics[width=.4\textwidth, page=1]{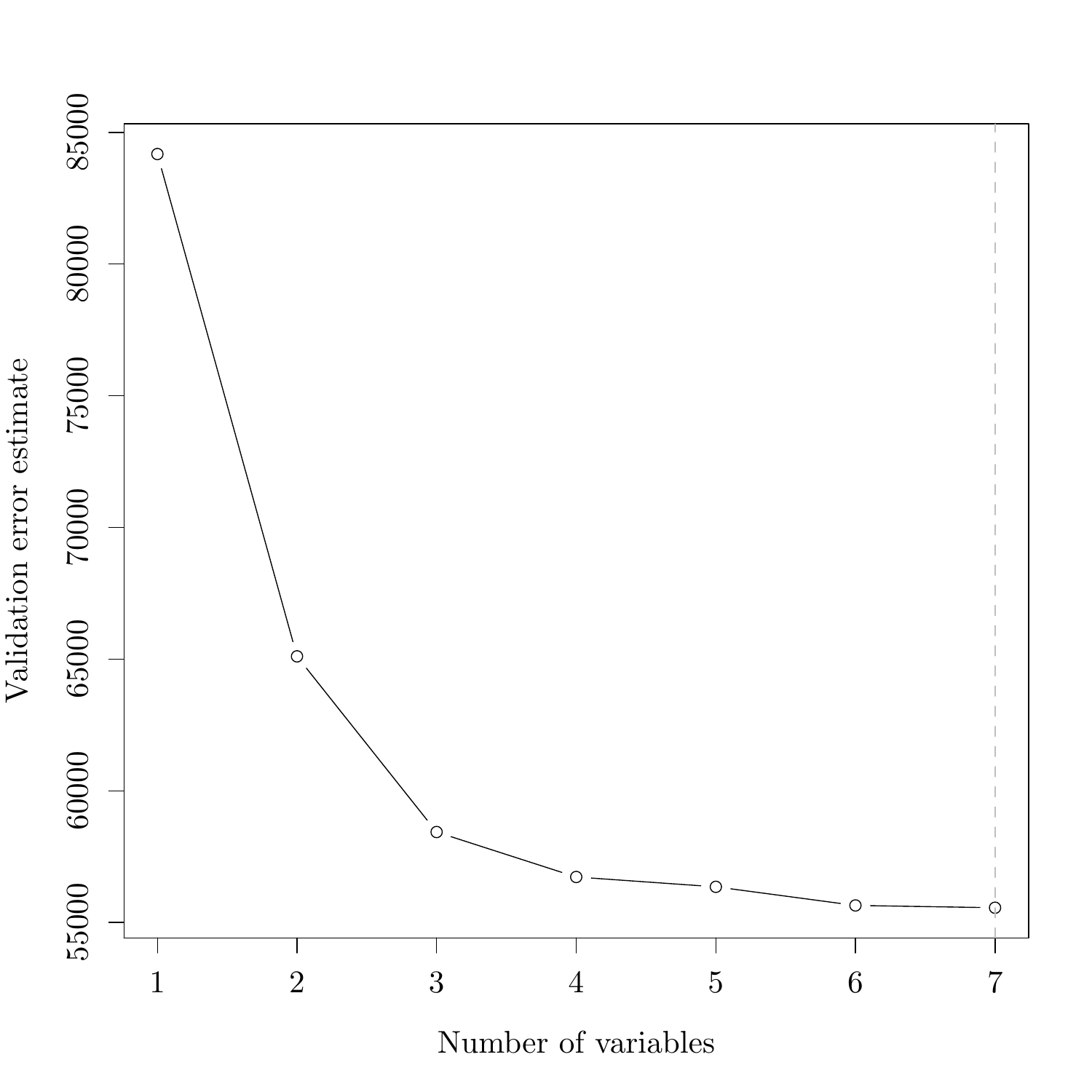}\label{fig:longLandingMSE_levAOA}}
	\subfloat[Angle of attack -- Selection frequencies]{\includegraphics[width=.4\textwidth, page=2]{levSel_longLandingEAP_AOA_tikz.pdf}\label{fig:longLandingfreq_levAOA}}
\caption{Application to  long landings -- selection of  wavelet levels for the gross weight and  angle of attack.}
	 \label{fig:longLanding_lev}
	\end{center}	
\end{figure}

We now determine, for a given flight parameter, which wavelet levels are the most able to predict the risk of long landing.
We perform selection of  wavelet levels  independently for the gross weight (GW.KG) and  angle of
attack (AOA), which are among the most commonly selected flight parameters.
Figures~\ref{fig:longLandingMSE_levGW} and \ref{fig:longLandingMSE_levAOA} show that the average number of
selected levels for GW.KG is less than for AOA. Indeed, the selection frequencies in
Figure~\ref{fig:longLandingfreq_levGW} indicate that for GW.KG, the first approximation levels are selected at each
run (groups $\zeta, G(0)$ and $G(1)$) whereas the last levels are selected in less than 40 of the 100 trials. The
situation is quite different for AOA: all  levels are selected more than 50 times. The
predictive power of this functional variable is contained in both the high levels of approximation and in the details of
the wavelet decomposition (Fig.~\ref{fig:longLandingfreq_levAOA}).

\subsubsection*{Detection of important time intervals}

\begin{figure}
	\begin{center}
	\subfloat[Altitude]{\includegraphics[width=0.4\textwidth, page=1]{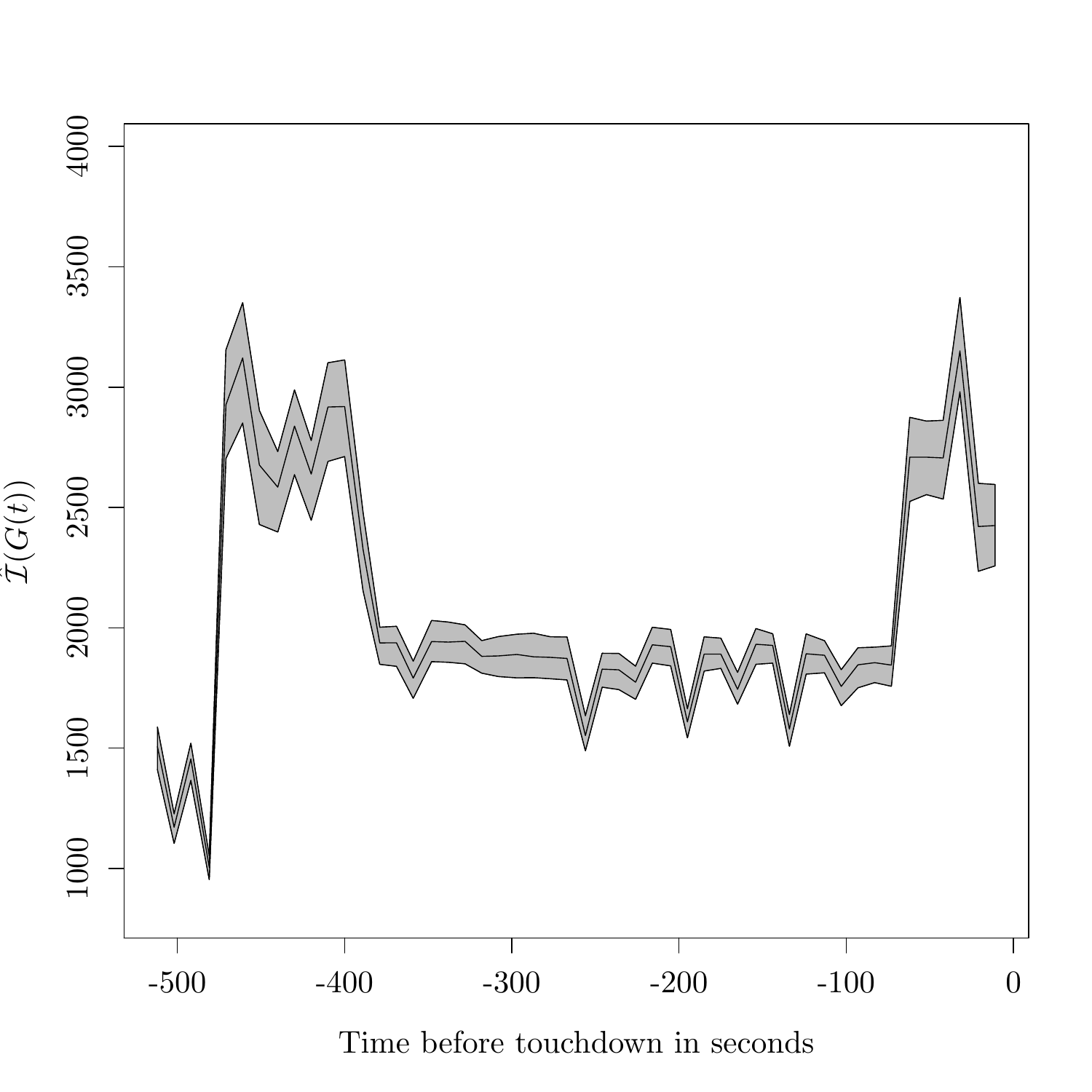}\label{fig:longLandingMSE_TimeSeqALT}}
	\subfloat[Angle of attack]{\includegraphics[width=0.4\textwidth, page=2]{impTemp_fdata.pdf}\label{fig:LevlongLandingfreq_TimeSeqAOA}}
	\caption{Application to  long landings -- averaged time importance and first and third quartiles over 100 trials.}
	\label{fig:longLanding_TimeSeq}
	\end{center}	
\end{figure}

The importance of time interval is now computed for two of the most important flight parameters, the altitude (ALT.STDC) and angle of attack (AOA).
Figure~\ref{fig:longLanding_TimeSeq} displays the averaged grouped importance $G(t)$ evaluated on 50 equally-spaced
time points. 
This number is arbitrarily chosen in order to find a tradeoff between computational time and accuracy.
The time $t=0$ refers to touchdown of the aircraft.

Two intervals are detected with high predictive power for the altitude. These results are
consistent with the view of aviation safety experts. Indeed, during the interval $[-460,-400]$, the
aircraft has to level off for stabilizing before the final approach. An overly high altitude at this moment can induce a
long landing. During the interval $[-60,0]$, the final seconds before touchdown, an overly high altitude can also induce a long
landing.

The interval detected for the angle of attack is  $[-200,-100]$. This make sense because according to flight procedure  pilots have to reduce the vertical speed a few seconds before touchdown (the flare). Consequently, they must increase the angle of attack in order to keep sufficient lift.

\section{Conclusion}

\begin{sloppypar}
We have considered the selection of grouped variables using random forests and proposed a new permutation-based importance measure for groups of variables. Our theoretical analysis provided exact decompositions of the grouped importance measure into a sum of the individual importances for specific models such as additive regression models. A simulation study highlighted the fact that in general the importance of a group does not reduce to the sum of the individual importances.
Since the idea of variable importance is not restricted to the random forest algorithm, it would be of interest to extend our method to other learning algorithms such as different bagging methods, neural networks and SVMs.

We have also introduced a rescaled version of the grouped variable importance with a heuristic choice for the normalization factor. The question of choosing a good value for the normalization factor is an open question that could perhaps be examined with further mathematical work.
\end{sloppypar}

The second contribution of the article was an original method for selecting functional variables using the grouped variable importance measure and a projection-based approach. The various functional variables were projected onto a wavelet basis and the corresponding coefficients  grouped in different ways. For instance, one can choose to regroup all coefficients of a given functional variable. The selection of these groups involved  selection of the functional variables. Various other groupings were proposed for wavelet decompositions. A backward algorithm adapted from recursive feature elimination in the context of random forests was used as a selection algorithm. The selection process was guided by the grouped importance measure in order to select the most relevant group of variables for predicting the variable of interest.

Lastly, an extensive simulation study was performed to illustrate  application of the grouped importance measure for functional data analysis, and encouraging results were then obtained for a real life problem in aviation safety. Of course, the grouped variable importance measure may be used for  any other application as soon as the features can be partitioned into known disjoint pieces. Potential applications can be found in the extensive literature about the group lasso method \citep{yuan2006model}, see for instance~\cite{meier2008group,ma2007supervised} for applications in Genetics and \cite{chatterjee2012sparse} for applications in  Climatology.

\appendix
\section{Proof of Proposition~\ref{prop:addModel} }
\label{sub:proof}

The vector $\mathbf X'_J$ and the vector  $\mathbf X_{(J)}$ are defined as in Section \ref{sec:RFgrimp}. We have
\begin{align*}
\mathcal I(\mathbf X_J) 	&= \E[(Y - f(\mathbf X) + f(\mathbf X) - f(\mathbf X_{(J)}))^2] - \E[(Y-f(\mathbf X))^2]\\
		&= \E[(f(\mathbf X) - f(\mathbf X_{(J)}))^2] + 2 \E \left[ \varepsilon (f(\mathbf X) - f(\mathbf X_{(J)})) \right] \\
		&= \E[(f(\mathbf X) - f(\mathbf X_{(J)}))^2],
\end{align*}
since $ \E [ \varepsilon  f(\mathbf X) ] = \E [   f(\mathbf X) \E[ \varepsilon | \mathbf X ] ]  = 0 $ and $ \E[ \varepsilon  f(\mathbf X_{(J)} ] = \E( \varepsilon)   \E [  f(\mathbf X_{(J)} ] = 0  $. Since $f(\mathbf X) = f_{J} (\mathbf X_{J})  +  f_{\bar J} (\mathbf X_{\bar J})$, we find that
\begin{align*}
\mathcal I(\mathbf X_J) 	&= \E[(f_J(\mathbf X_J) - f_J(\mathbf X'_J))^2]\\
		&= 2\Var[f_J(\mathbf X_J)],
\end{align*}
as $\mathbf X_J$ and  $\mathbf X'_J$ are independent and identically distributed.

\section{Additional experiments on the grouped variable importance}
\label{sub:ExpPIGV}

In this section, we investigate the properties of the permutation importance measure of groups of variables with
numerical experiments, in addition to the theoretical results given before. In particular, we compare this quantity with
the  sum of individual importances  in various models. We also study how this quantity behaves in ``sparse situations''
where only a small number of variables in the group are relevant for  predicting the outcome.

The general framework of the experiments is the following. For a fixed $p \geq 1$, define $\mathbf X^\top  := (\mathbf
W^\top, \mathbf Z ^\top)$ where $\mathbf W$ and $\mathbf Z$ are  random vectors both of length $p$. Some of the
components of $\mathbf W$ are correlated with $Y$ whereas those in $\mathbf Z$ are all independent of
$Y$.  Let  $\textbf{C}_w$ be the variance-covariance matrix of $\mathbf W$. By incorporating the group $\mathbf Z$ in the model,
we present a realistic framework  where not all the $X_j$ have a  link with $Y$. For each experiment, we simulate
$n=1000$ samples of $Y$ and $\mathbf X$ and   compute the importance  $\mathcal I (\mathbf W)$ of the group $\mathbf
W$, the rescaled grouped variable importance  $\mathcal I_{\textrm{nor}} (\mathbf W)$ and the sum of the individual importances of
the variables in $\mathbf W$. We repeat each experiment 500 times. The boxplots of the importances over the 500
repetitions are shown in Figures \ref{fig1a1b}  to \ref{fig1c1d} with values $p$
between 1 and 16.

Let $\textbf{0}_p$ and $\textbf{I}_p$ denote the null vector and  identity matrix of  $\R^p $.
Let $\indd_p$ be the vector in $\R^p$ with all coordinates equal to one and let
$\textbf{0}_{p,q} $  denote the null matrix of dimension $p \times q$.

\subsubsection*{Experiment 1: linear link function.}

We simulate $\mathbf X$ and $Y$ from a multivariate Gaussian distribution. More
precisely, we simulate
samples from  the joint distribution
$$
\begin{pmatrix}
\mathbf X \\
Y
\end{pmatrix}
 =
\begin{pmatrix}
\mathbf W \\
\mathbf Z \\
Y
\end{pmatrix}
\sim \mathcal N _{2p+1} \left(\textbf{0}_{2p+1}\, , \,
\begin{pmatrix}
	  	\textbf{C}_w & \textbf{0}_{p,p} &  \boldsymbol \tau \\
		 \textbf{0}_{p,p} & \textbf{I}_{p} & \textbf{0}_p  \\
		 \boldsymbol \tau^\top&  \textbf{0}_p^\top &  1
\end{pmatrix} \right) ,
 $$
 where $\boldsymbol \tau$ is the vector of the covariances between $\mathbf W$ and $Y$. In
this context, the conditional distribution of $Y$ over $\mathbf X$ is normal and
the
conditional mean $f$  is a linear function: $f(\mathbf x) = \sum_{j=1}^{p+q} \alpha_j
x_j$ with $\alpha = (\alpha_1, \ldots, \alpha_p, 0 , \dots, 0 )^\top$ a sequence of
deterministic coefficients (see for instance \citet{rao73}, p.~522.

\begin{itemize}
\item{\bf  Experiment 1a: independent predictors}. We take $\boldsymbol
\tau = 0.9 \, \indd_{p}$ and $\textbf{C}_w = \textbf{I}_p$. All  variables of $\mathbf W$ are
independent and correlated with $Y$.
\item{\bf  Experiment 1b: correlated predictors}. We take $\boldsymbol \tau
= 0.9 \, \indd_{p}$ and $\textbf{C}_w = (1-0.9) \textbf{I}_{p} + 0.9 \indd_p \indd_p^\top$. The
variables of $\mathbf W$ are correlated, and also correlated with
$Y$.
\item{\bf Experiment 1c: independent predictors, sparse case}. We take
$\boldsymbol \tau = (0.9, 0 ,
\dots, 0)^\top$ and $\textbf{C}_w =  \textbf{I}_p$. Only the first variable in the group $\mathbf W$ is correlated with $Y$.
\item{\bf Experiment 1d: correlated predictors, sparse case}. We take
$\boldsymbol \tau = (0.9, 0, \dots, 0)^\top$ and  $\textbf{C}_w = (1-0.9) \textbf{I}_{p} + 0.9 \indd_p \indd_p^\top$. The
variables of $\mathbf W$ are correlated. Only the first variable in the group $\mathbf W$ is correlated with $Y$.
\end{itemize}

\subsubsection*{Experiment 2: additive link function.}
We simulate $\mathbf X$ from a multivariate Gaussian distribution:
$$
\mathbf W =
\begin{pmatrix}
\mathbf W \\
\mathbf   Z \\
\end{pmatrix}
\sim \mathcal N _{2p} \left( \textbf{0}_{2p}\, , \, \begin{pmatrix}
	  	\textbf{C}_w & \textbf{0}_{p,p} \\
		 \textbf{0}_{p,p} & \textbf{I}_{p}
\end{pmatrix} \right) , $$
and the conditional distribution of $Y$ is
 $$(Y | \mathbf X)  \sim  \mathcal N  \left(    \sum_{j=1}    ^p      f_j(X_j),
1   \right) , $$
where $f_j(x) = \sin(2x) + j$  for $j<p/2$ and $f_j(x) = \cos(2x) + j$  for $j
\geq p/2$.

\begin{itemize}
\item {\bf Experiment 2a: independent predictors}. We take  $\textbf{C}_w = \textbf{I}_p$. All
variables of $\mathbf W$ are independent and correlated with $Y$.
\item {\bf Experiment 2b: correlated predictors}.  We take $\textbf{C}_w = (1-0.9) \textbf{I}_{p}
+ 0.9 \indd_p \indd_p^\top$. The variables of $\mathbf W$ are correlated and also correlated with $Y$.
\end{itemize}

\subsubsection*{Experiment 3: link function with interactions.}
We simulate $\mathbf X$ from a multivariate Gaussian distribution:
$$
\mathbf W =
\begin{pmatrix}
\mathbf W \\
\mathbf   Z \\
\end{pmatrix}
\sim \mathcal N _{2p} \left( \textbf{0}_{2p}\,  \textbf{I}_{2p}  \right) , $$
and the conditional distribution of $Y$ is
 $$(Y | \mathbf X)  \sim  \mathcal N  \left( \sum_{j=1} ^p  X_j + X_p  X_1 +  \sum_{j=1} ^{p-1}  X_j X_{j+1}
 \,   , \,  1   \right) .$$

\begin{figure}
	\begin{center}
	\subfloat[Experiment 1a]{\includegraphics[width=0.4\textwidth]{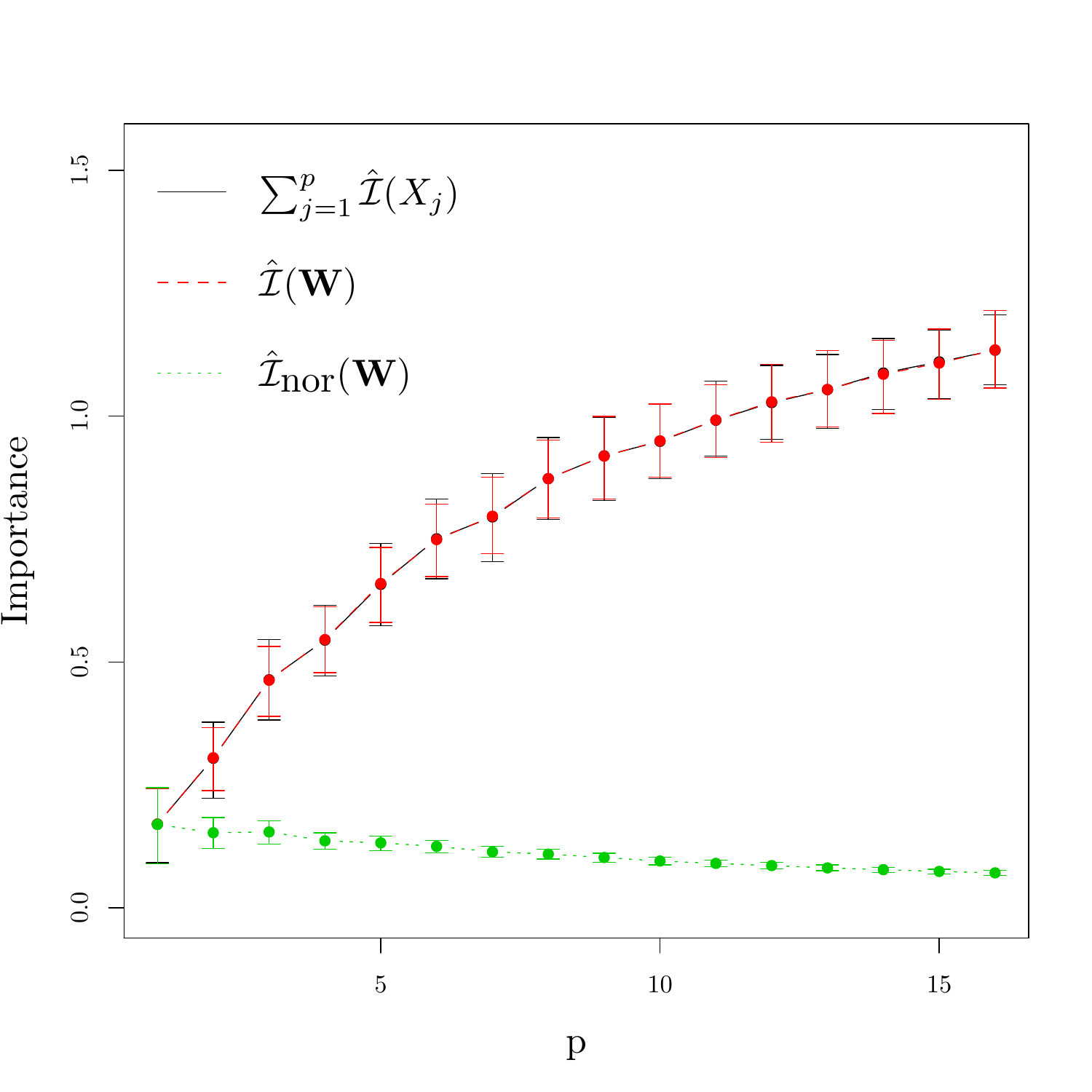}}
	\subfloat[Experiment 1b]{\includegraphics[width=0.4\textwidth]{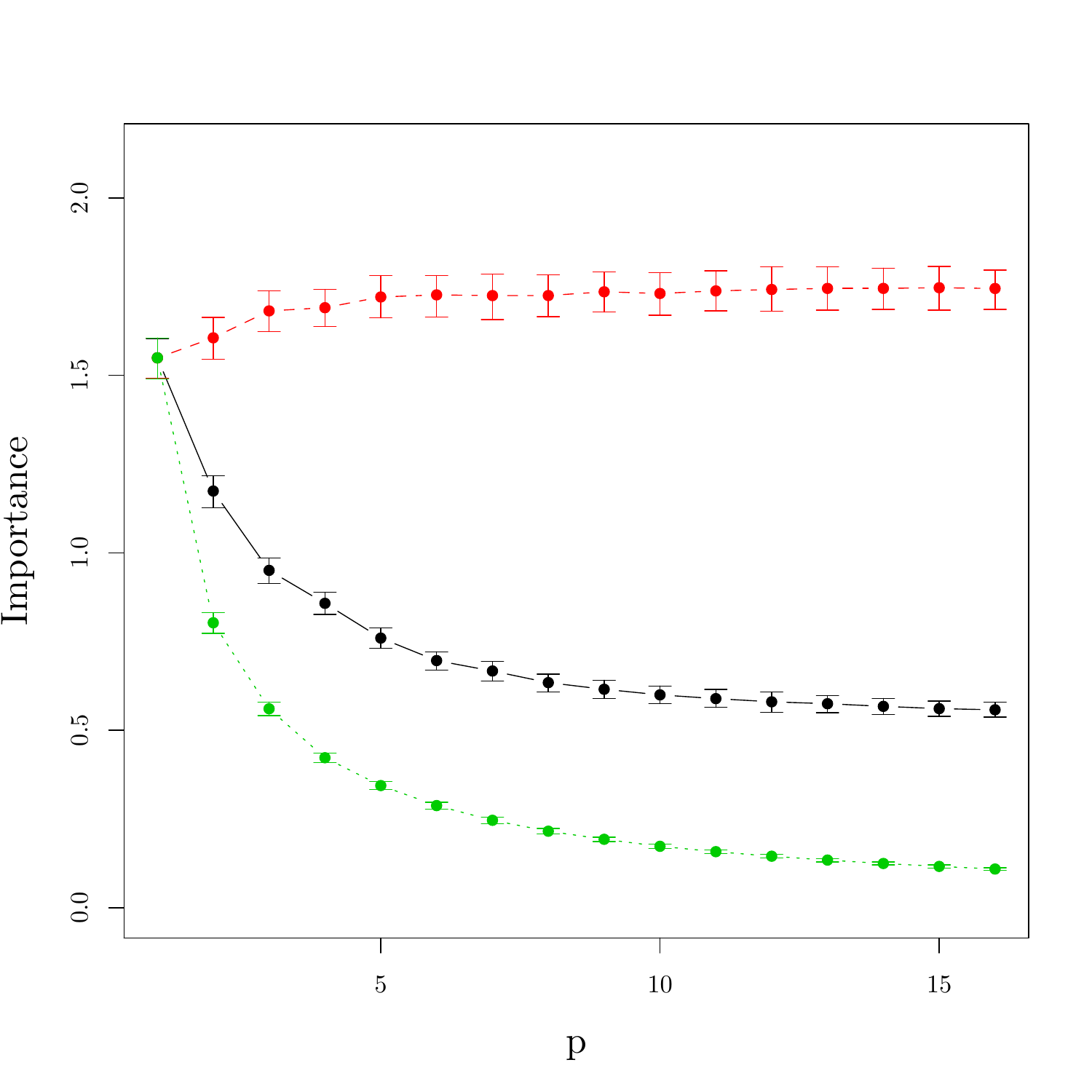}}
	\caption{Boxplots of the importance measures for Experiments 1a and 1b.
The number of variables in $\mathbf W$ varies from 1 to 16. For Experiment 1a,
the sum  of the individual importance@s and $\hat { \mathcal I }(\mathbf W)$ overlap.}
	 \label{fig1a1b}	
	\end{center}	
\end{figure}

\begin{figure}
	\begin{center}
	\subfloat[Experiments 2a]{\includegraphics[width=0.4\textwidth]{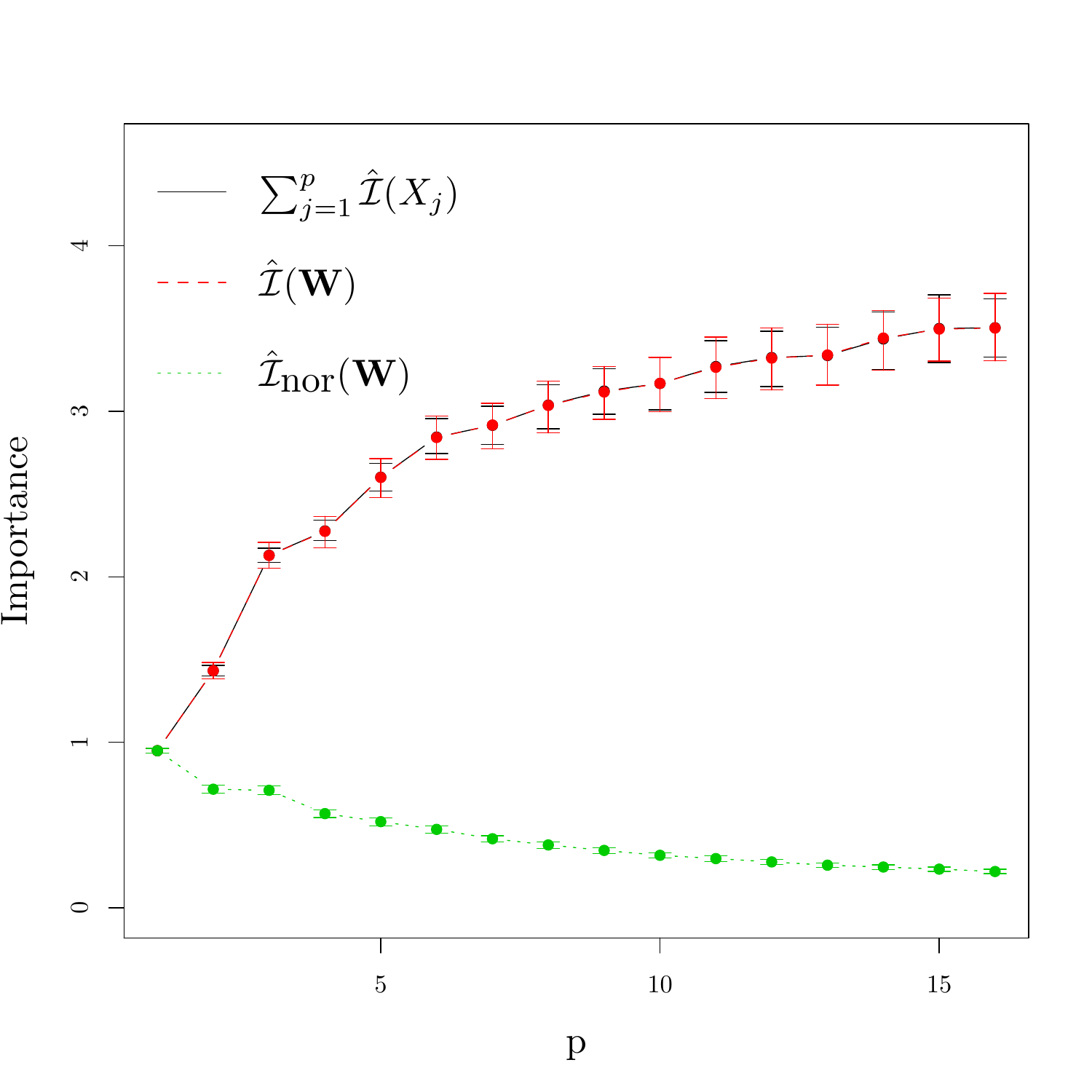}}
	\subfloat[Experiments 2b]{\includegraphics[width=0.4\textwidth]{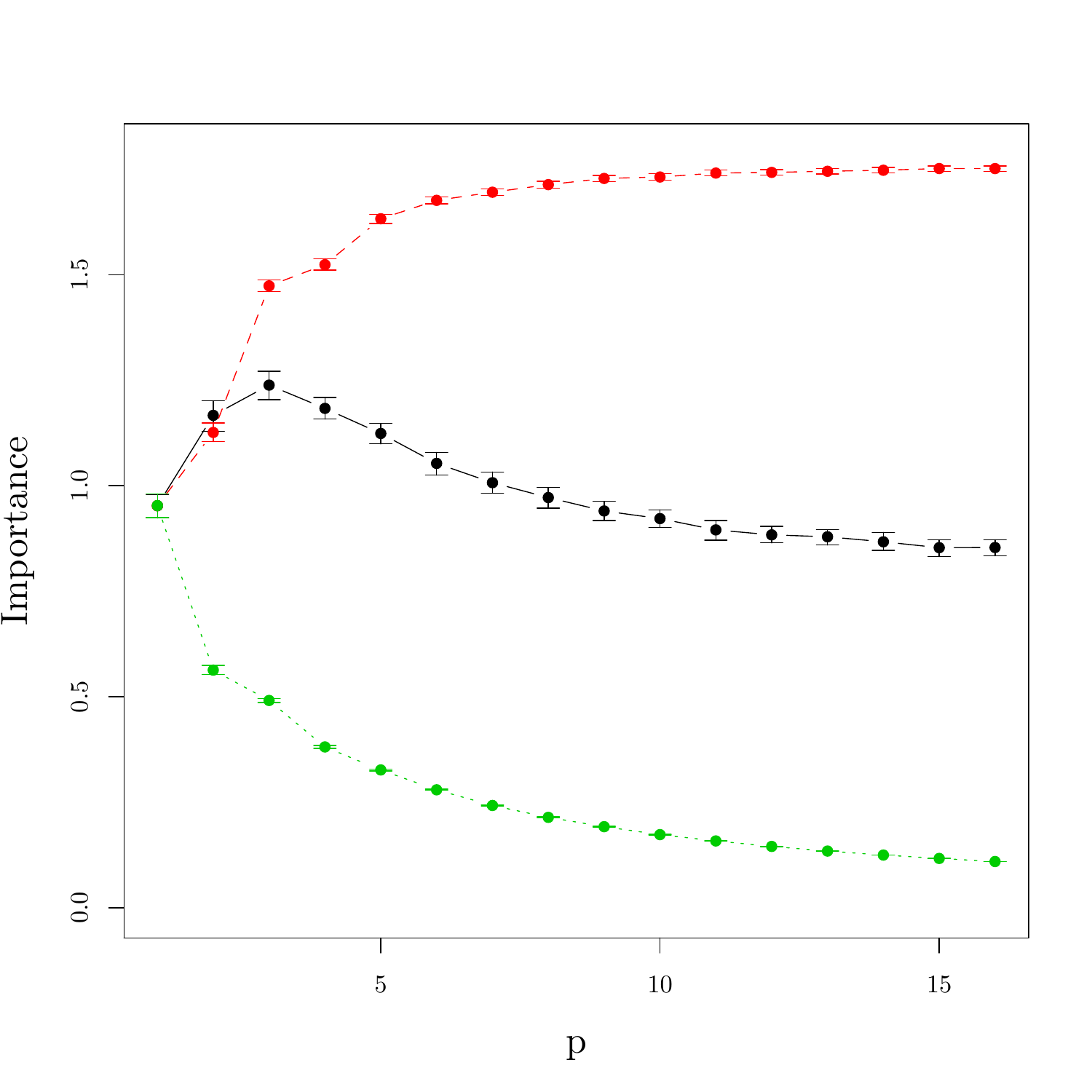}}
	\caption{Boxplots of the importance measures for Experiments 2a and 2b.
The number of variables in $\mathbf W$ varies from 1 to 16. For Experiment 2a, the sum  of the individual importances
and $\hat { \mathcal I } (\mathbf W)$ overlap.}	 	
	 \label{fig2a2b}		
	\end{center}
\end{figure}

\begin{figure}
	\begin{center}
	\includegraphics[width=0.4\textwidth]{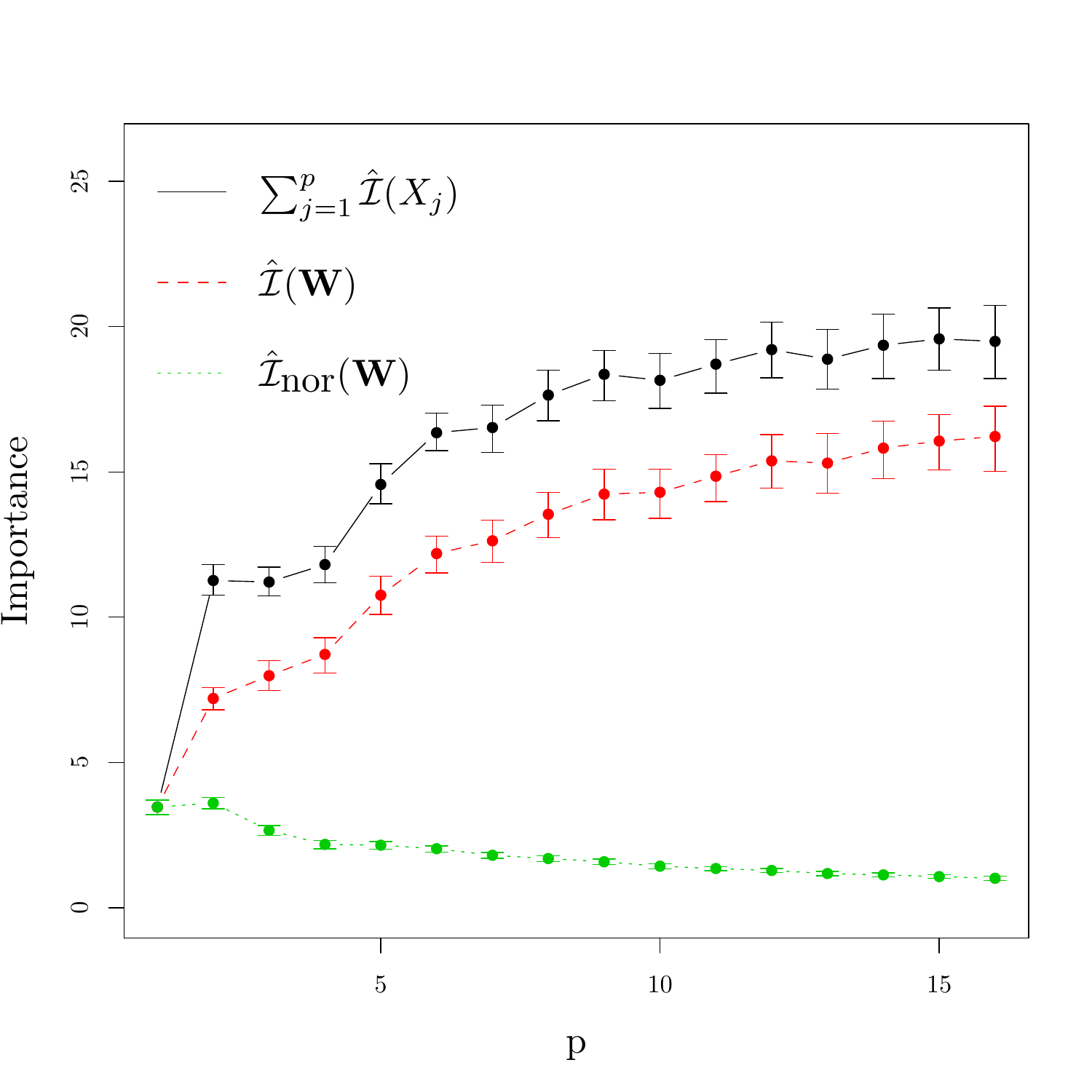}
	\caption{Boxplots of the importance measures for Experiment 3.
The number of variables in $\mathbf W$ varies from 1 to 16. }	 	
	 \label{fig3}		
	\end{center}
\end{figure}

\begin{figure}
	\begin{center}
	\subfloat[Experiment 1c]{\includegraphics[width=0.4\textwidth]{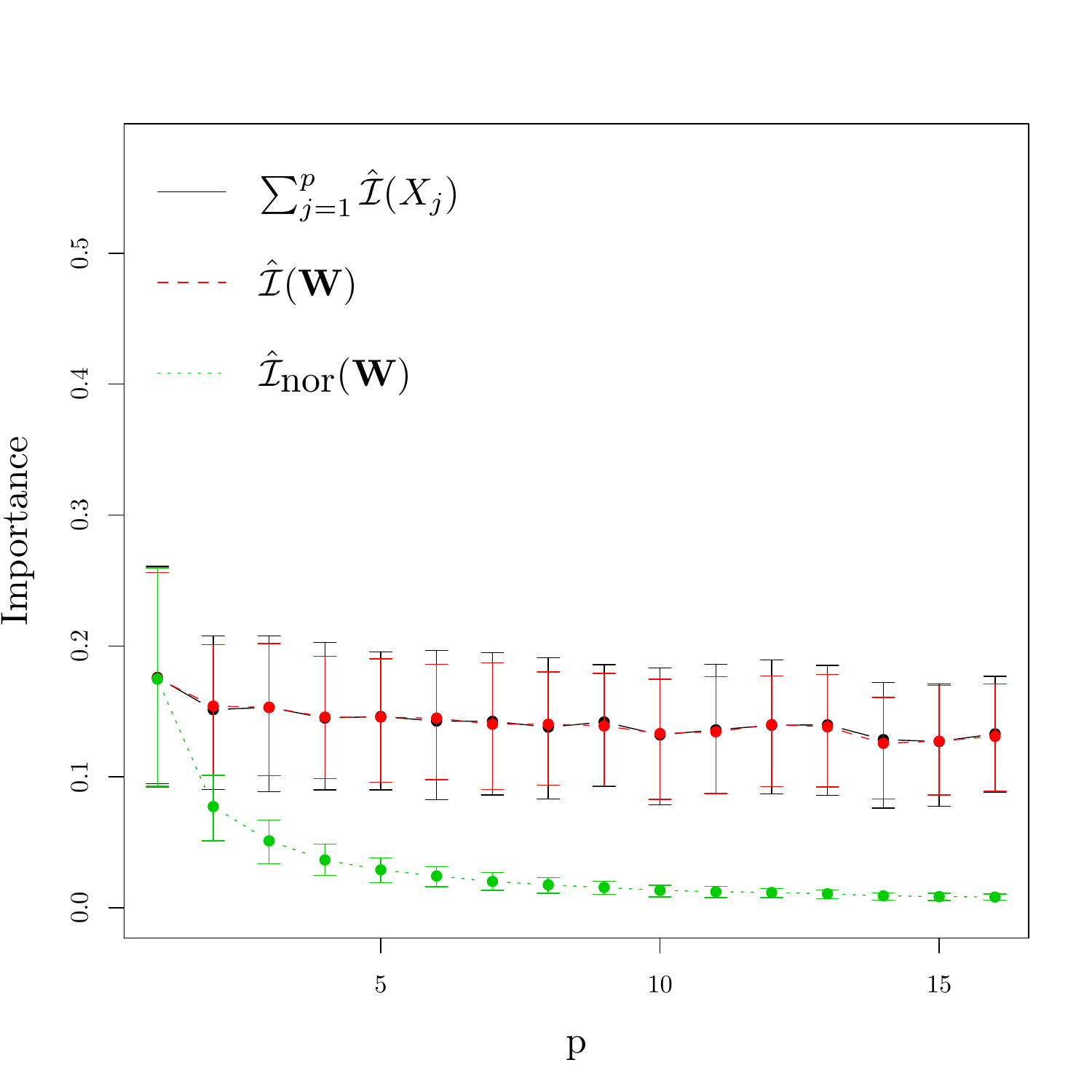}}
	\subfloat[Experiment 1d]{\includegraphics[width=0.4\textwidth]{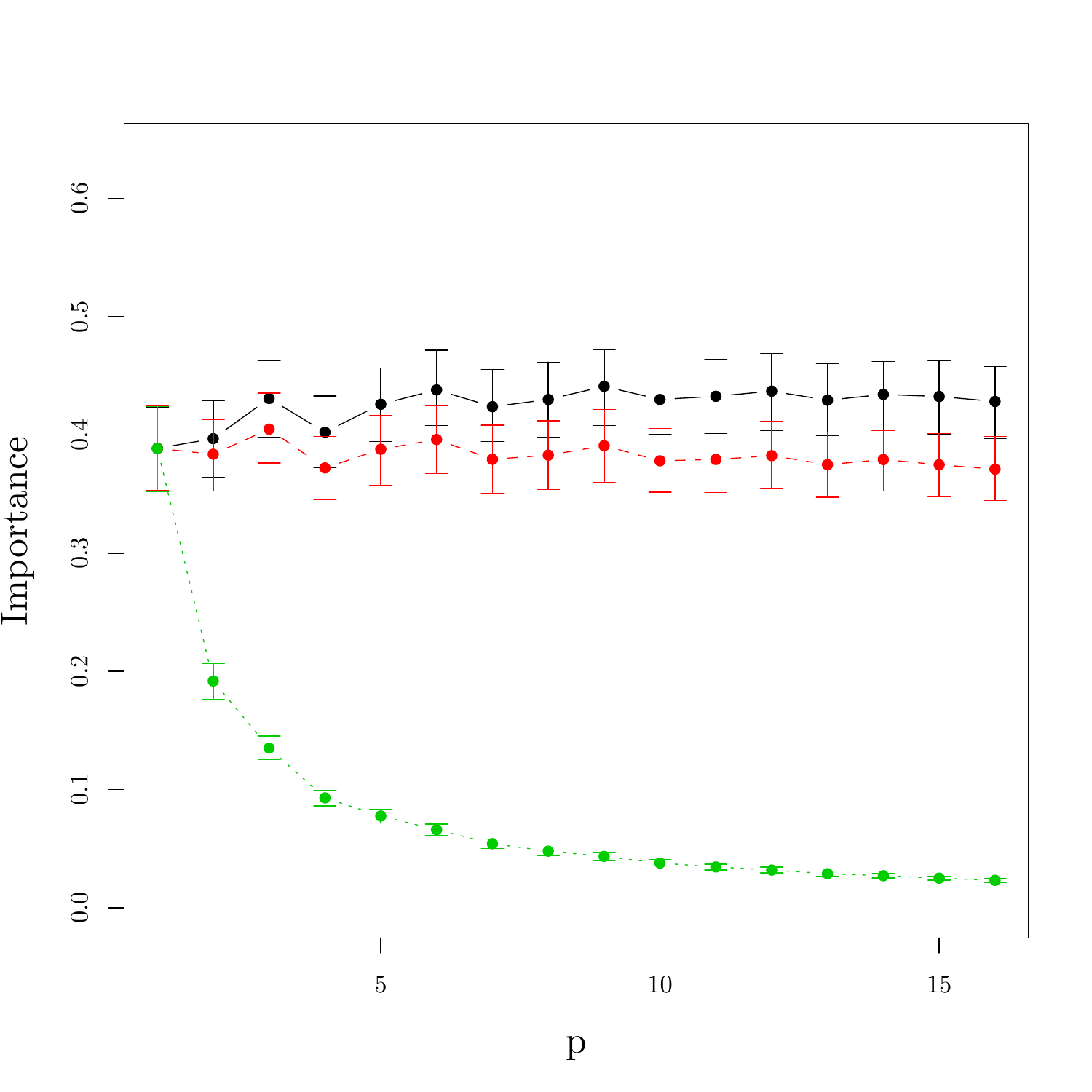}}
	\caption{Boxplots of the importance measures for Experiments 1c and 1d.
The number of variables in $\mathbf W$ varies from 1 to 16. For Experiment 2a,  the sum  of the individual importances
and $\hat { \mathcal I } (\mathbf W)$ overlap.}
		 \label{fig1c1d}
	\end{center}
\end{figure}

\subsubsection*{Results}
Experiments 1a-b and 2a-b illustrate the results of Corollary~\ref{cor:specificmodels} (see Figures~\ref{fig1a1b} and \ref{fig2a2b}). Indeed, the regression function in both cases satisfies the additive property
(\ref{cond:addtit}) for these experiments. In Experiments 1a and 2a, the variables of the group $\mathbf W$ are
independent and the grouped variable importance is nothing more than the sum of the individual importances in this case.
In Experiments 1b  and 2b, the variables of  group $\mathbf W$ are positively correlated. In these
situations, the grouped variable importance is larger than the sum of the individual importances, which agrees with
Equation \ref{ImportanceCorrel}. Note that the grouped variable importance increases with $p$, which is natural because
the amount of information for predicting $Y$ increases with the group size in these models. On the other hand, it was shown
in \cite{rf:G+14} that individual importances decrease with correlation between the predictors. Indeed we observe that
the sum of the individual importances decreases with $p$ in the correlated cases.

The regression function in Experiment 3 does not satisfy the additive form \eqref{cond:addtit}. Although the variables
in the group are independent, the grouped variable importance is not equal to the sum of the individual importances (Figure~\ref{fig3}). In a general setting therefore, it appears that these two quantities differ.

We now comment the results of the sparse Experiments 1c and 1d (Figure~\ref{fig1c1d}).  It is clear that
$ \mathcal I ( \mathbf  W ) = \mathcal I (X_{1})  = \sum_{j  = 1 \dots p} \mathcal I (X_{j})  $ for these two
experiments (see Proposition~\ref{prop:addModel} for instance). Regarding Experiment 1c, the boxplots of the estimated
values $ \hat{\mathcal I} ( \mathbf  W ) $ and $ \sum_{j  = 1 \dots p}  \hat{ \mathcal I}  (X_{j}) $ agree with this
equality. On the other hand, in Experiment 1d, we observe that  $  \sum_{j  = 1 \dots p} \hat{ \mathcal I}  (X_{j}) $ is
significantly higher than $ \hat{\mathcal I} ( \mathbf  W ) $. Indeed, it has been noticed by  \cite{rf:N+10} that the
individual importances of predictors that are not associated with the outcome tend to be overestimated  by  random
forests when there correlation between predictors. In contrast,
 the estimator $ \hat{\mathcal I} ( \mathbf  W ) $ seems to correctly estimate $ \mathcal I ( \mathbf  W ) $ even
for large $p$.  Indeed, for both experiments 1c and 1d, the importance $ \hat{\mathcal I} ( \mathbf  W )$ is unchanged
when the size of the group $p$ varies from 2 to 16. Note that for variable selection, we may prefer to consider the
rescaled importance $\hat{\mathcal I} _{\textrm{nor}}$ so as to select with priority small groups of variables.

\section{Curve dimension reduction with wavelets}
\label{sec:dimRed}

The analysis of flight data in Section~\ref{sec:appli} required, for computational reasons, to preliminarily reduce the
dimension of the wavelet decomposition of the flight parameters. We require to adapt the well-known wavelet shrinkage method to
the context of independent random processes. Using the notation of Section~\ref{sec:contrib}, we first recall the
hard-thresholding estimator introduced by \citet{wave:DJ94} in the case of one random signal. This approach is then
extended to deal with $n$ independent random signals.

\subsection*{Signal denoising via wavelet shrinkage}

The problem of signal denoising can be summarized as follows. Suppose that we observe $N$ noisy samples $X(t_1), \dots,
X(t_N)$ of a deterministic function $s$ (the signal):
\begin{equation}
X(t_\ell) = s(t_\ell) + \sigma \varepsilon_\ell, \quad \ell=1,\ldots,N ,
\label{eq:signalModel}
\end{equation}
where the  $\varepsilon_{\ell}$ are  independent standard Gaussian random variables. We assume that $s$ belongs to $L^2([0,1])$.
The goal is to recover the underlying function $s$ from the noisy data $\{X(t_\ell), \ell \in \{1,\dots,N\}\}$  with small error.
Using the discrete wavelet transform, this model can be rewritten in the wavelet domain as
$$\xi_{jk} = \omega_{jk} + \sigma \eta_{jk}, \quad  \forall j \in \{0, \dots, J-1\}, \, \forall k \in \{ 0, \dots,
2^j-1\},$$
and the scaling domain as
$$\zeta = \omega_{0} + \sigma \eta_{0},$$
where $\xi_{jk}$ and $\zeta$ are the empirical wavelet and scaling coefficients of $X(t_\ell)$ as in
Equation~\eqref{eq:waveRepresTrunc}.
The random variables $\eta_{jk}$ and $\eta_{0}$ are i.i.d. random variables from the distribution $\mathcal N_1(0,1)$.

A natural approach for estimating $\omega_{jk}$ is to shrink the coefficients $\xi_{jk}$ to zero. An estimator of $s$ in
this context has the form
\begin{equation}
\hat s(t_\ell) = \hat \omega_{0} \phi(t_\ell) + \sum_{j=0}^{J-1} \sum_{k=0}^{2^j-1} \hat \omega_{jk} \psi_{jk}(t_\ell) ,
\label{eq:signalEst}
\end{equation}
with $\hat \omega_0 = \zeta$ and
\begin{equation*}
\hat \omega_{jk} = \left\lbrace
\begin{array}{lll}
\xi_{jk} & \mbox{ if } |\xi_{jk}| > \delta_N\\
0 & \mbox{ otherwise.}
\end{array}
\right.
\end{equation*}
This method is referred to as the {\it hard-thresholding estimator} in the literature. \citet{wave:DJ94} propose the universal
threshold $\delta_N=\sigma \sqrt{2 \log(N)}$.
In addition, the standard deviation $\sigma$ can be estimated by the median absolute deviation (MAD)
estimate of the wavelet coefficients at the finest levels, i.e.,
$$\hat \sigma = \dfrac{\mbox{Med}(|\xi_{jk} - \mbox{Med}(\xi_{jk})|: j=J-1, k=0, \dots, 2^{J-1}-1)}{0.6745},$$
where the normalisation factor 0.6745 comes from the normality assumption in \eqref{eq:signalModel}. This estimator is
known to be a robust and consistent estimator of $\sigma$. The underlying idea is that the variability in the wavelet
coefficients is essentially concentrated at the finest level.

\subsection*{Consistent wavelet thresholding for independent random signals}

\paragraph{$\bullet$ Identically distributed case}

We start by assuming that the observations come from the same distribution: for any $i \in  \{1,\dots , n \}$,
\begin{equation}
\label{eq:multicurvesIID}
X_i(t_\ell) = s(t_\ell) + \sigma \varepsilon_{i,\ell}, \quad \ell=1,\ldots,N,
\end{equation}
where the $\varepsilon_{i,\ell}$ are independent standard Gaussian random variables. The wavelet coefficients of $s$ can be easily deduced from the mean signal $\bar X := \frac 1 n \sum_{i=1} ^n X_i$ which satisfies
 \begin{equation*}
\bar X (t_\ell) = s(t_\ell) + \frac{\sigma}{\sqrt n } \varepsilon_{\ell}, \quad \ell=1,\ldots,N,
\end{equation*}
where the $\varepsilon_{\ell}$ are independent standard Gaussian random variables. By applying the hard-thresholding rule to this signal, we obtain the following estimation of the wavelet parameters of $s$:  $\hat  \omega_0 = \zeta$ and
\begin{equation*}
\hat \omega_{jk} = \left\lbrace
\begin{array}{lll}
\bar \xi_{jk} & \mbox{ if } | \bar \xi_{jk}| > \bar \delta_N\\
0 & \mbox{ otherwise,}
\end{array}
\right.
\end{equation*}
where $\bar \xi_{jk}$ is the wavelet coefficient of level $(j,k)$ of $\bar X $. Here the threshold is $\bar \delta_N= \frac {\sigma} {\sqrt n} \sqrt{2 \log(N)}$.

\paragraph{$\bullet$ Non identically distributed case}

In many real life situations, assuming that the $n$ signals are identically distributed is not a realistic
assumption. For the study presented in Section~\ref{sec:appli} for instance, the flight parameters have no reason to follow the same distribution in safe and unsafe conditions. We propose a generalization of
the model \eqref{eq:multicurvesIID} by introducing a latent random variable $Z$ taking its values in a set $\mathcal Z$. Roughly speaking, the variable $Z$
represents all the phenomena that have an effect on the mean signal.
 Conditionally on $Z_i=z_i$, the
distribution of the process $X_i$ is now defined, for any $i \in \{1,\dots,n\}$, by
\begin{equation}
\label{eq:multicurves}
X_i(t_\ell) = s(t_\ell,z_i) + \sigma \varepsilon_{i,\ell} \, ,  \quad \ell=1,\ldots,N,
\end{equation}
where the $\varepsilon_{i,\ell}$ are independent standard Gaussian random variables. This regression model allows us
to
consider various situations of interest
arising in functional data analysis. In  supervised
settings where a variable $Y$ has to be predicted using $X$, one reasonable  model is to  take $Z=Y$. We now propose
a hard-thresholding method which simultaneously shrinks the wavelet decomposition of
the $n$ signals.

Let $\Vert \cdot \Vert_n$ denote the $\ell_2$-norm in $\R^n$: $\Vert\boldsymbol u \Vert_n := \sqrt{\sum_{i=1}^n
u_{i}^2}$ for any $\boldsymbol u \in \R^n$.  Let $\boldsymbol{\xi}_{jk}$ be the vector
$(\xi_{1jk},\dots,\xi_{ijk},\dots, \xi_{njk})^\top$ where $\xi_{ijk}$ is the  coefficient   of level $(j,k)$ in the
wavelet decomposition of the signal $X_i$.
For any $z \in \mathcal Z$, let $\omega_{jk} (z)$ be the wavelet coefficient of level  $(j,k)$ of $s(\cdot,z)$, and $\boldsymbol \omega_{jk} := (\omega_{jk}(Z_1),\dots, \omega_{jk}(Z_n))^\top$. We define the common wavelet support of $s$ by
$$ L := \left\{ (j,k) \, | \,  \omega_{jk}(Z) = 0  \: a.s. \right\} .  $$
 If $(j,k) \in L$, then $\boldsymbol \omega_{jk} = (0,\dots, 0)^\top $  almost surely and  $\Vert\boldsymbol \xi_{jk}\Vert_n^2$ has a centered chi-square distribution with $n$ degrees of
freedom. Otherwise, $\boldsymbol \omega_{jk}$ can be non null and in this case $\Vert\boldsymbol \xi_{jk}\Vert_n^2$ has
the distribution
of a sum of $n$  independent uncentered chi-square distributions. We can
thus propose a thresholding rule for the statistic $\Vert\boldsymbol \xi_{jk}\Vert_n$. For any $ j \in  \{0, \dots,
J-1\}$ and any
$ k \in  \{ 0, \dots, 2^j-1\}$, let
\begin{equation}
\hat{\boldsymbol \omega}_{jk} = \left\lbrace
\begin{array}{ll}
\boldsymbol \xi_{jk} \, & \mbox{ if } \Vert\boldsymbol \xi_{jk}\Vert_n > \delta_{N,n}\\
(0,\dots,0)^\top \, & \mbox{ otherwise,}
\end{array}
\right.
\label{eq:mht}
\end{equation}
where the threshold $\delta_{N,n}$ depends on $N, n$ and $\sigma$.
Proving adaptive results in the spirit of  \cite{wave:D+95} for this method is beyond the scope of the paper. However,
an elementary consistent result can be proved.
We would like $\hat{\boldsymbol \omega}_{jk}$ to
be a zero vector with high probability when $(j,k) \in L$. For some $x \geq 0$, take $\delta^2 _{N,n} =
\delta_{N,n}^2(x) = \sigma ^2 ( 2x + 2 \sqrt{n x} + n)$. Then,
\begin{eqnarray}
\Proba \left[  \bigcup_{(j,k) \in L }   \,  \left\{  \hat{\boldsymbol \omega}_{jk}  \neq  (0,\dots,0)^\top  \right\}
\right]
&\leq& \sum_{(j,k) \in L } \Proba \left[ \Vert\boldsymbol \xi_{jk}\Vert^2_n \geq  \delta^2_{N,n}(x) \right] \notag  \\
&=& \sum_{(j,k) \in L } \Proba \left[ \dfrac{\Vert\boldsymbol \xi_{jk}\Vert_n^2}{\sigma^2} - n \geq 2x + 2 \sqrt{n x} \right]
\notag \\
&\leq& |\bar L|  e^{-x} \leq  N e^{-x} ,  \label{upperbound}
\end{eqnarray}
where we have used a deviation bound for central chi-square distributions from \citet[p.~1325]{ms:LM00}.  If the
signal is exactly zero, it can be recovered with
high probability by taking  $x \gg \log(N)$. In particular, if we choose $x = 2\log(N)$, the threshold is
$\delta_{N,n}^2 =  \sigma^2 (4\log(N) + 2 \sqrt{2 n \log(N)} + n)$ and the convergence rate in (\ref{upperbound})  is of
order  $O(\frac{1}{N})$. In practice,   $\sigma$ can be estimated by a MAD estimator computed on the coefficients of the
highest level of all  $n$ wavelet decompositions. Next, $x$ and $\delta_{N,n}$ can be chosen such that
(\ref{upperbound}) is lower than a
given probability $q$. Letting $Ne^{-x} = q$, we obtain the threshold $$\delta_{N,n} = \hat \sigma
\left(2\log\left(\dfrac{N}{q}\right) + 2 \sqrt{n \log\left(\dfrac{N}{q}\right)} + n \right)^{\frac{1}{2}}.$$

Assuming that $ \omega_{j,k}(Z) = 0$ almost surely for some level $(j,k)$ is a strong assumption that is difficult to meet
in practice. It can be hoped that this method still works if the wavelet support of $s(\cdot,z)$ does not vary too much with $z$.
In particular, it may be applied if there exists a common set $S$ of
indices $(j,k)$ such that, for any $z$, the projection of $s(\cdot,z)$  on $\mbox{Vect}(\psi_{jk} \, | \, (j,k) \in S )$
is not too far from $s(\cdot,z)$ for the $L^2$ norm.

\section*{References}

\bibliography{biblio}

\end{document}